\documentclass{article} 

\let\savedaddcontentsline\addcontentsline
\usepackage{iclr2027_conference,times}
\let\addcontentsline\savedaddcontentsline

\makeatletter
\renewcommand\iclrruler[1]{}
\makeatother

\usepackage{graphicx}
\usepackage{booktabs}
\usepackage{placeins}
\usepackage{amsmath,amssymb}
\usepackage{amsthm}

\newtheorem{proposition}{Proposition}

\usepackage{hyperref}
\hypersetup{hidelinks}
\usepackage{url}
\iclrfinalcopy

\makeatletter
\newcommand{\equalcontrib}{\textsuperscript{\textdagger}}
\newcommand{\corresponding}{\textsuperscript{\textasteriskcentered}}
\newcommand{\affiliations}[1]{\gdef\@affiliations{\ignorespaces#1}}
\def\@affiliations{}
\let\savedauthor\author
\renewcommand{\author}[1]{%
  \savedauthor{#1\ifx\@affiliations\@empty\else\\[1.6ex]\mdseries\@affiliations\fi}}
\makeatother

\begin{document}

\title{Policy-Level Recursive Self-Improvement for Embodied AI with a \mbox{Criticality World Model}}

\author{
    \normalsize\bf
    Linxuan He,\textsuperscript{\rm 1,2}\equalcontrib
    Yuying Tian,\textsuperscript{\rm 1,2}\equalcontrib
    Lingxiang Fan,\textsuperscript{\rm 1,2}
    Jiaqi Pi,\textsuperscript{\rm 1}\\[0.9ex]
    \bf Yinqiao Lu,\textsuperscript{\rm 1}
    Shang Su,\textsuperscript{\rm 1,2}
    Mengkai Shi,\textsuperscript{\rm 2}
    Shuo Feng\textsuperscript{\rm 1}\corresponding
}
\affiliations{
    \smallskip
    \textsuperscript{\rm 1}Tsinghua University \\[0.7ex]
    \textsuperscript{\rm 2}Dense-AI \\[0.7ex]
    fshuo@mail.tsinghua.edu.cn \\[0.7ex]
    \small\textdagger Equal contribution. \textasteriskcentered Corresponding author.
}

\maketitle

\lhead{}
\renewcommand{\headrulewidth}{0pt}

\begin{abstract}
Despite advances in policy pretraining, embodied AI systems can plateau during task-specific fine-tuning as uniform scenario collection encounters fewer of the remaining failures. We address this problem with a policy-level recursive self-improvement loop: execution outcomes train a criticality world model, whose risk scores guide scenario collection for specialist training. To account for the resulting shift in scenario frequencies, weighted resampling approximately corrects the collection bias. Because specialized updates can weaken nominal behavior, a risk-based gate selects between the specialist and a frozen nominal policy at inference. We extend this procedure to reinforcement learning and behavior cloning. A sampling analysis characterizes the correction, while a controlled study examines the trade-off between critical and nominal coverage. Across five embodied domains, the composed systems reduce failure rates by 59--67\% for locomotion and manipulation and 8--25\% for VLA benchmarks relative to their respective baselines.
\end{abstract}

\section{Introduction}

Pretrained policies for embodied AI have advanced rapidly, with vision-language-action (VLA) models and foundation models demonstrating broad cross-task capabilities \citep{physicalintelligence2025pi06,selfimproving2025neurips}. Yet task-specific fine-tuning can plateau below the success rates needed for dependable operation. A policy may perform well in nominal conditions while further training brings only marginal gains.

This stagnation is not only a capacity problem but also a data problem \citep{chen2025revisiting}. As a policy improves, uniformly sampled perturbations yield fewer remaining failures within a fixed rollout budget. This is the \emph{curse of rarity} \citep{liu2024curse,feng2023dense}. In embodied AI, the search spans high-dimensional perturbations including action noise, external forces, object poses, and visual conditions \citep{tao2025maniskill3,liu2023libero,reddi2024qarl}. State-dependent proposals help find rare failures \citep{delecki2025failure}, while replay methods distinguish uniform and prioritized batch sampling \citep{mnih2015human,schaul2016prioritized}.

\begin{figure}[t]
\centering
\includegraphics[width=0.96\textwidth]{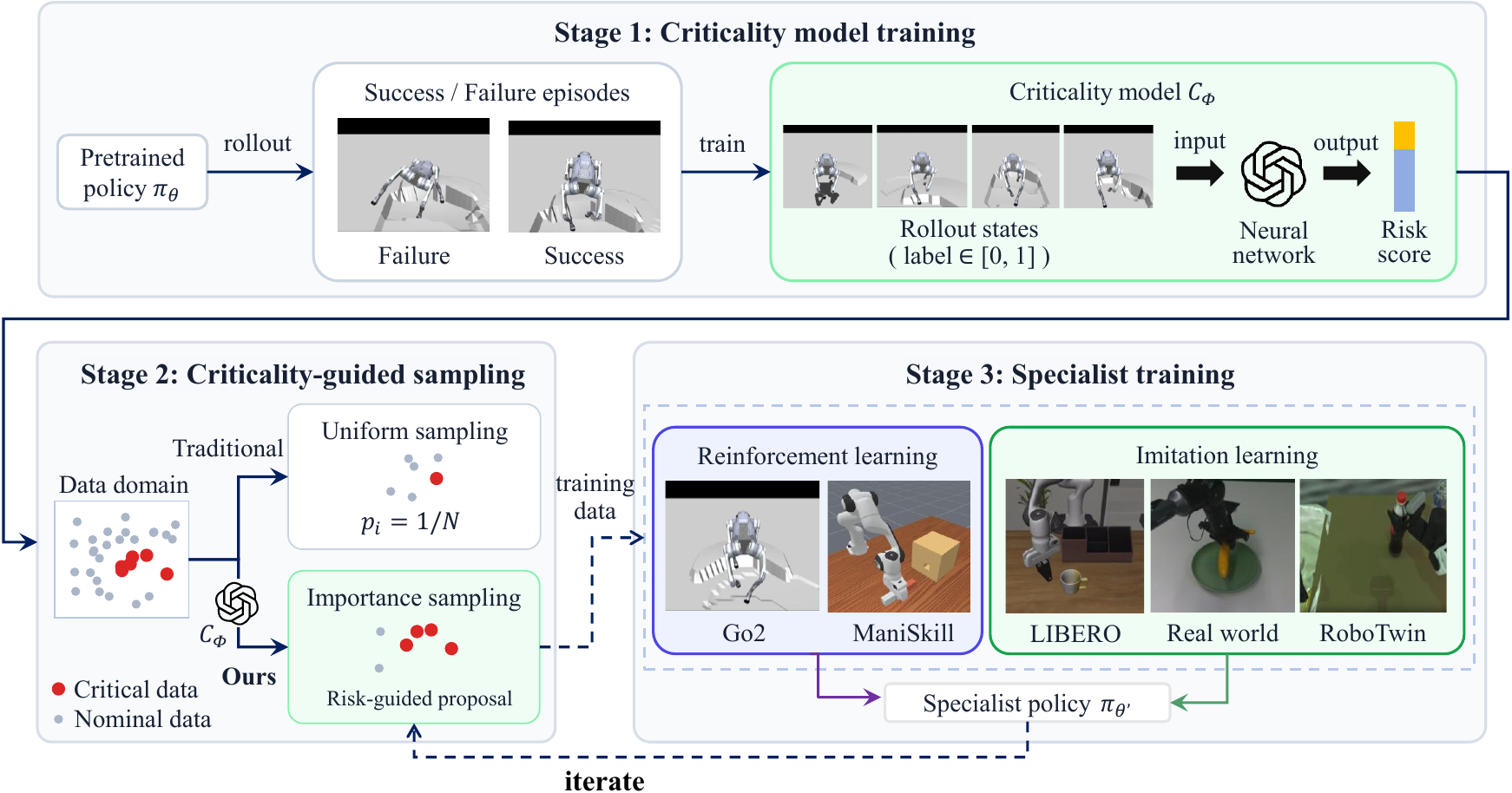}
\caption{Training loop. Rollout outcomes train the criticality model, whose risk score guides specialist-data collection. New rollouts can repeat the update. The diagram illustrates enrichment of critical scenarios. The positive-support proposals and subsequent weighted resampling are defined in Section~\ref{sec:method}.}
\label{fig:framework}
\end{figure}

Dense learning addresses this challenge in autonomous driving through informative-data selection and a safety specialist with risk-based takeover \citep{feng2026breaking}. We extend this foundation across robot systems and learning paradigms. Our policy-level recursive self-improvement (RSI) loop trains a lightweight criticality world model $C_\phi$ (hereafter, criticality model) from execution outcomes and uses its score to guide specialist-data collection (Figure~\ref{fig:framework}). It predicts the risk of a future task failure without generating trajectories; its output is a ranking score rather than a calibrated probability. Repeated rounds evaluate the updated composition while retaining the learning procedure.

Our central learning mechanism is specialist training guided by the criticality world model. Concentrating training on critical scenarios can, however, weaken nominal behavior through reduced coverage or optimization interference. We therefore complement training with a risk-based gate that selects between the specialist and a frozen nominal policy. To assess gains beyond this composition, we also gate a policy fine-tuned on randomly collected data (Random + gate).

The extension has two parts. For reinforcement learning (RL), we instantiate the loop for quadruped locomotion under terrain perturbations and manipulation under external forces, adapting outcome labels, candidate perturbations, and policy-update interfaces to each system. For behavior cloning (BC), rollout outcomes determine which expert demonstrations to select, while expert actions supply policy supervision. Because risk-guided collection changes scenario frequencies, weighted resampling approximately corrects this shift while retaining each domain's update objective. Section~\ref{sec:theory} states the exact importance-sampling identity and its finite-buffer limits; the correction uses scenario probabilities without requiring action densities.

We evaluate Go2 locomotion in MuJoCo \citep{todorov2012mujoco}, ManiSkill3 manipulation \citep{tao2025maniskill3}, LIBERO \citep{liu2023libero}, RoboTwin~2.0 \citep{chen2025robotwin2}, and AgileX Piper manipulation. The imitation-learning comparisons match starting checkpoints, data quantities, and optimization settings. ManiSkill provides a same-round comparison without selective gating. Go2 separates policy updates from gate calibration, with different budgets and hyperparameters across arms. We report shared-round and final results.

Our primary contributions are
\begin{enumerate}
\item We extend criticality-guided improvement from driving to RL-pretrained locomotion and manipulation, with terrain- and force-selection interfaces, task-specific failure labels, and multi-round updates. These implementations test the approach across different robot dynamics and training objectives.
\item We extend the workflow to BC and VLA policies by separating rollout-based failure labels from expert-action supervision. Inverse-proposal resampling retains existing supervised objectives; the analysis specifies the corrected demonstration-pool objective and its finite-buffer limits.
\item Across five domains, we document a same-round ManiSkill gain without selective gating and lower failure point estimates than Random + gate in three imitation-learning domains. Component and fixed-gate comparisons, Go2 calibration results, and a controlled coverage study clarify these findings.
\end{enumerate}

\section{Related Work}

\paragraph{Dense learning and cross-system extension.} Closest to our work, \citet{feng2026breaking} derive a gradient-informed training distribution and combine an RL driving specialist with risk-based takeover. Their method identifies avoidable crashes and near-misses, densifies informative states, and reevaluates data through counterfactual simulation. Building on this foundation, we extend its empirical scope in two directions. First, Go2 and ManiSkill require terrain- and force-conditioned proposals, task-specific failure labels, and different RL policy updates. Second, our BC extension separates rollout-based risk evidence from expert-action supervision through finite-pool resampling compatible with existing VLA losses. Our contribution therefore lies in these implementations and cross-system findings, building on established importance sampling and risk-based takeover.

\paragraph{Rare-event sampling and policy repair.} More broadly, NADE \citep{feng2021intelligent}, APE \citep{xu2022ape}, and state-dependent proposals \citep{delecki2025failure} target rare-event estimation; our proposal also allocates training data. At the buffer level, prioritized replay reweights stored transitions \citep{schaul2016prioritized}, whereas our proposal changes which scenarios enter. For policy repair, RADIUM offers gradient-accelerated and gradient-free sampling \citep{dawson2025radium}. We instead score finite candidate sets with a learned outcome predictor. For expert-data acquisition, ADC uses human perturbations and corrective teleoperation \citep{huang2025adc}; our BC implementation selects scenes through predicted risk.

\paragraph{Risk prediction, rewards, and intervention.} Beyond collection, SAFE predicts VLA failures from internal features \citep{gu2026safe}. Our risk scores guide both collection and routing, although our LIBERO and RoboTwin gates additionally require simulator state. For reward learning, other methods convert failures or demonstrations into dense rewards \citep{wu2025robust,tsao2026svm,koprulu2025dense}; our model supplies proposal and routing scores. Human supervision also supports SERL and HIL-SERL through demonstrations and, in HIL-SERL, corrections \citep{luo2024serl,luo2025hilserl}. To allocate such supervision, ThriftyDAgger and Fleet-DAgger schedule human interventions \citep{hoque2021thriftydagger,hoque2023fleetdagger}. Our deployment gate instead selects between learned policies; BC receives expert supervision during data collection.

\paragraph{Iterative embodied learning.} $\pi^\star_{0.6}$ learns from on-robot experience and corrections \citep{physicalintelligence2025pi06}; LiReN continually adapts a navigation policy \citep{stachowicz2025liren}. Related systems use autonomous practice, search, or reward-policy coevolution \citep{selfimproving2025neurips,seea2025neurips,huang2025roska}. For VLA improvement, PLD trains residual RL specialists and distills their rollouts into a generalist \citep{xiao2026pld}. In comparison, our BC route selects expert demonstrations and retains both policies at deployment. This preserves a fallback but incurs input, memory, and inference costs. Our evaluation tests multiple rounds in Go2 and ManiSkill and one update in each imitation-learning domain.

\paragraph{Scope of the predictive model.} Finally, self-referential RSI can modify its learning procedure \citep{schmidhuber2003goedel}; our update rule stays fixed. On the modeling side, generative world models predict future states \citep{ha2018worldmodels}, while value-equivalent models retain decision-relevant predictions \citep{grimm2020value} and visual world models predict robot-action outcomes \citep{liu2024siriusfleet}. Within this spectrum, our criticality world model predicts only future failure risk, supporting the different scenario proposals and training objectives evaluated here.

\section{Sampling Analysis and Coverage Motivation}
\label{sec:theory}

\subsection{Risk-Informative Scenarios}

Let $\mu_\theta(s)$ be the failure probability of policy $\pi_\theta$ conditional
on an initial scenario $s$. Its population risk is $\mu_\theta=\mathbb{E}_{s\sim P}[\mu_\theta(s)]$. This conditional failure probability is a population quantity, distinct from the learned risk score $C_\phi$. For this risk objective, define the scenarios with nonzero conditional gradients as
\begin{equation}
\Phi_{\rm in}=\{s:\|\nabla_\theta\mu_\theta(s)\|_2>0\}.
\label{eq:phi_in}
\end{equation}
If $P(\Phi_{\rm in})$ is small, uniform collection encounters few such scenarios. Low aggregate failure alone does not imply this condition. Moreover, successful demonstrations can have nonzero BC gradients outside this set. The learned risk score's association with gradient support remains an empirical assumption.

\subsection{Proposal and Finite-Buffer Correction}

Following the gradient-informed proposal in \citet{feng2026breaking}, let $P(s)$ be the target distribution over a finite scenario pool and $g_\theta(s)$ the conditional gradient of the chosen training objective. A one-sample importance-weighted gradient has trace variance $\sum_s P(s)^2\|g_\theta(s)\|_2^2/q(s)-\|\mathbb E_Pg_\theta\|_2^2$. Without a support floor, the minimizing proposal is
\begin{equation}
q^\star(s)\propto P(s)\|g_\theta(s)\|_2.
\label{eq:qstar}
\end{equation}
For the BC implementation, let $D_i$ be the expert demonstration associated with candidate scene $i$ and let $g_{\rm BC}(i)=\nabla_\theta\ell_{\rm BC}(\theta;D_i)$ for its supervised action loss. Conditional on a fixed demonstration pool, the objective-specific gradient and its exact proposal correction satisfy
\begin{equation}
G_{\rm BC}=\sum_i p(i)g_{\rm BC}(i)
=\mathbb E_{I\sim q}\!\left[\frac{p(I)}{q(I)}g_{\rm BC}(I)\right].
\label{eq:bc_change_measure}
\end{equation}
For fixed per-demonstration gradients, Equation~\ref{eq:qstar} also applies to BC with $P=p$ and $g_\theta=g_{\rm BC}$. Its ideal proposal depends on the supervised action-loss gradient rather than the failure-risk gradient. If a demonstration loss includes random within-trajectory draws, $g_{\rm BC}(i)$ denotes their conditional expected gradient, while the variance-optimal proposal for the resulting stochastic estimator depends on its second moment.
In practice, however, failure risk need not rank gradient magnitude. Some failures are irreducible, while successful near-misses may have large behavior-cloning gradients. We therefore use $C_\phi$ to enrich the candidate pool for potentially avoidable failures, not as an exact implementation of Equation~\ref{eq:qstar}. Binary cross-entropy on rollout outcomes trains the score, and class rebalancing or shaped labels make it a ranking signal rather than a calibrated probability. The variance calculation describes the ideal proposal, for which the score is an empirical proxy.

Although the score is only a proxy, positive support wherever $P(s)>0$ allows the usual change of measure for a fixed integrand $f$.
\begin{equation}
\mathbb E_{s\sim q}[w(s)f(s)]=\mathbb E_{s\sim P}[f(s)],\qquad w(s)=P(s)/q(s).
\label{eq:unbiased}
\end{equation}
If a candidate pool is independently sampled from a broader scenario distribution, an unnormalized $p/q$ estimator is unbiased for that broader expectation after averaging over both stages. In training, we instead normalize ratios within a realized selected buffer and resample batches. This approximately restores the finite-pool objective but is not exactly unbiased at finite buffer size. For evaluation, we use the unnormalized estimator where proposal probabilities are recorded (Appendix~\ref{app:is_eval}).

Appendix~\ref{app:theory_proofs} gives the two-stage proof.

\subsection{Distinct Critical Examples}

Beyond correcting sampling frequencies, selection also changes which failure modes a finite buffer covers. The following calculation describes the effect of gradient redundancy, conditional on its assumptions.
\begin{proposition}
\label{prop:rho}
For a fixed gradient coordinate or projection and $N\geq2$, let $Z,Z_1,\ldots,Z_N$ have common variance $\sigma^2>0$. Pool~A repeats the same random draw $Z$ $N$ times. Pool~B contains $Z_1,\ldots,Z_N$ with average pairwise correlation $\bar\rho$. Over repeated pool construction, the ratio of their sample-mean variances is
\begin{equation}
\frac{\operatorname{Var}_{B}}{\operatorname{Var}_{A}}=\frac{1+(N-1)\bar\rho}{N}.
\label{eq:rho}
\end{equation}
\end{proposition}
Independent gradients give a ratio of $1/N$; perfectly correlated ones give $1$. This motivates distinct critical examples when they lower gradient correlation, an unmeasured property of our robot-policy gradients. Appendix~\ref{app:theory_proofs} gives the proof.

Under a fixed training budget, broader critical coverage can reduce nominal coverage. If $N_{\rm nom}$ distinct nominal gradients are independent with variance $\sigma_{\rm nom}^2$, their sample-mean variance is $\sigma_{\rm nom}^2/N_{\rm nom}$. Resampling reuses examples but cannot add nominal scenarios to a finite pool. Consequently, a specialist may improve in selected failure regions while becoming less reliable elsewhere. Retaining a frozen $\pi_{\rm nom}$ preserves its original behavior, although the composed policy depends on selecting it appropriately. The controlled study below illustrates this coverage trade-off; it does not run the learned proposal or measure robot-policy gradient correlations.

\paragraph{Controlled coverage test.} We examine scenario coverage in a 2-D
binary classification task with a dominant nominal rule and ten Gaussian
``exception'' regions near its decision boundary. Each region denotes a
distribution of examples, not one fixed point. The two training pools have the
same 110-example batch size and the same 10:100 critical-to-nominal ratio.
Pool~A repeats one critical region and covers ten nominal regions. Pool~B
covers ten critical regions and repeats one nominal region. This exchange
changes both empirical distributions, so the comparison tests a coverage
trade-off rather than an importance-weighted estimate of one common target.

\begin{figure}[t]
\centering
\includegraphics[width=0.78\textwidth]{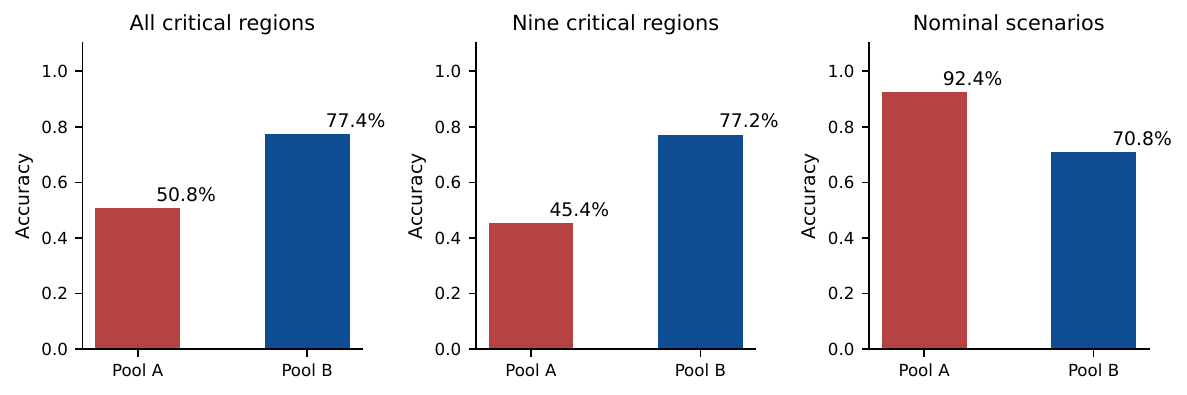}
\caption{Controlled coverage result averaged over ten seeds. Pool~B improves accuracy on critical regions unseen by
Pool~A but gives up nominal accuracy. The pools have equal batch size and
critical-to-nominal sample ratio.}
\label{fig:main_results}
\end{figure}

Across ten seeds, the same two-hidden-layer multilayer perceptron (MLP) trains for 500 epochs. On the nine critical regions absent from Pool~A, Pool~B reaches 77.2\% accuracy versus 45.4\% for Pool~A. Conversely, nominal accuracy is 70.8\% versus 92.4\% (Figure~\ref{fig:main_results}). Appendix~\ref{app:toy} gives the design and learning curves.

\section{Method}
\label{sec:method}

\paragraph{Criticality model.} Each domain supplies a candidate scenario or perturbation, an outcome label, and a policy-update objective. We roll out the current policy and pair its available context with the eventual outcome. RL contexts combine per-step observations and candidate terrain or force perturbations; BC contexts describe initial scenes linked to expert demonstrations. Thus unsuccessful policy rollouts train the predictor, while expert actions supervise the BC policy. A lightweight MLP learns from context--label pairs $(z,y)$ by minimizing binary cross-entropy:
\begin{equation}
\mathcal L_{\rm crit}(\phi)=-\mathbb E_{(z,y)}\!\left[y\log C_\phi(z)+(1-y)\log\bigl(1-C_\phi(z)\bigr)\right].
\label{eq:criticality_loss}
\end{equation}
Here $y=1$ denotes a positive failure label under the domain's labeling rule. Inputs, labels, and architectures are given in Appendices~\ref{app:go2}--\ref{app:criticality_summary}.

Labels come from the policy being improved. Go2 down-samples safe timesteps and labels the last steps before a failure as positive. ManiSkill associates each observation--force pair with its episode outcome. The three imitation-learning domains use initial-state episode outcomes. The score therefore ranks risk within each collection setup rather than estimating a calibrated failure probability. In the multi-round experiments, new rollouts supply labels after each specialist update.

\paragraph{Scenario proposal.} Using this score, we define a proposal over a finite candidate set $\mathcal S$:
\begin{equation}
q(s)=(1-\varepsilon)\frac{\kappa(s)}{\sum_{s'}\kappa(s')}+\frac{\varepsilon}{|\mathcal S|},\qquad 0<\varepsilon\leq1.
\label{eq:q_rl}
\end{equation}
Go2 uses $\kappa=C_\phi$ with an activation cutoff of 0.3. ManiSkill uses $\kappa=\exp(3C_\phi)$ when its maximum candidate score exceeds 0.5, and samples uniformly otherwise. These collection cutoffs differ from the deployment gate threshold. For per-step perturbations $u_t$ given rollout history $h_t$, the recorded ratio is $w_t=P(u_t\mid h_t)/q(u_t\mid h_t)$ and the trajectory ratio is $W=\prod_t w_t$, provided the remaining transition mechanisms are unchanged. For an imitation-learning demonstration pool, the corresponding proposal and ratio are
\begin{equation}
q(i)=\varepsilon p(i)+(1-\varepsilon)\frac{\kappa(s_0^{(i)})}{\sum_j\kappa(s_0^{(j)})},\qquad w(i)=p(i)/q(i).
\label{eq:q_il}
\end{equation}
Here $p(i)$ is the target mass of demonstration $i$ in the candidate pool, uniform when each candidate has equal mass. The ratio corrects scenario selection, not action likelihood. Task stratification or replay changes that target and must be accounted for separately.

\begin{figure}[t]
\centering
\includegraphics[width=0.88\textwidth]{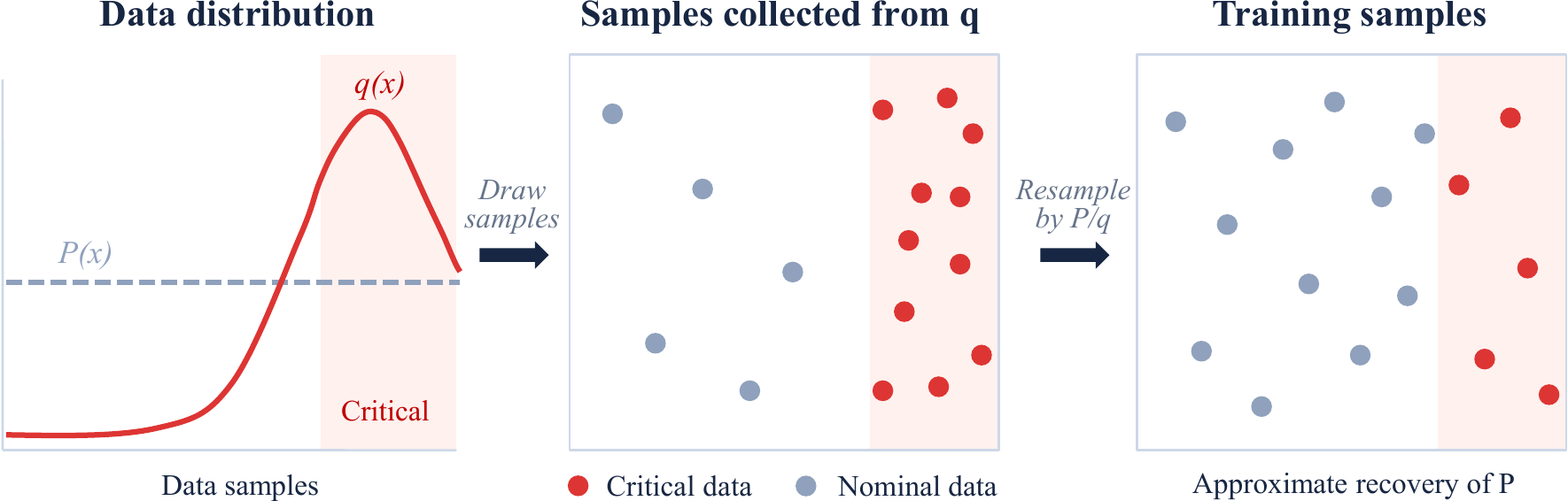}
\caption{Criticality-guided selection and finite-buffer correction. The
proposal $q$ (red) enriches critical scenarios relative to target $P$ (blue).
Resampling selected examples by normalized $P/q$ ratios approximates the
target using only the realized buffer. Markers schematically indicate sample
frequencies. Resampling cannot supply unselected scenarios.}
\label{fig:importance_resampling}
\end{figure}

Figure~\ref{fig:importance_resampling} separates risk-guided collection from
the subsequent training draw. Resampling changes the frequencies of scenarios
already in the buffer but cannot recover scenarios that were never selected.
The positive-support mixture prevents zero proposal probabilities, although
concentrated proposals can still produce large correction weights.

Candidate sets contain $10^4$ terrain cells for Go2, $11^3$ force vectors for ManiSkill, demonstration pools for LIBERO and RoboTwin, and $10^4$ physical-task scenarios. The categorical $P$ and $q$ remain computable for policies without tractable action densities. Concentrated proposals and products of per-step ratios can produce large weights, motivating the uncertainty analysis in Appendix~\ref{app:is_eval}.

\paragraph{Specialist update.} For the RL-pretrained policies, normalized trajectory-ratio resampling feeds domain-specific updates. Go2 uses filtered, advantage-weighted regression with a BC anchor and failure penalties, related to advantage-weighted actor-critic learning \citep{nair2020awac}. ManiSkill uses offline proximal policy optimization (PPO) with a BC anchor and a value loss. In LIBERO, RoboTwin, and the real-robot task, expert demonstrations are selected by $q(i)$. BC batches are then resampled from the realized subset in proportion to their recorded $w(i)$. Repeated selections retain their sampling mass. These finite-buffer procedures approximate correction of the collection distribution. Collection and updating repeat for six rounds in Go2 and 17 in ManiSkill, while the other three experiments use one specialist update.

For these BC updates, the loss remains unweighted: outcomes train the predictor, expert actions supervise the policy, and inverse-proposal ratios determine batch frequencies. Because the ratios use candidate-scene probabilities without an action-likelihood term $\pi_\theta(a\mid s)$, flow-based and diffusion-based VLA policies retain their supervised objectives. The experiments compare guided selection plus resampling against uniformly collected data.

Despite this sampling correction, the update losses differ from the binary failure-risk objective analyzed in Section~\ref{sec:theory}. The correction addresses only the scenario proposal. It does not undo return filtering, advantage weighting, or additional regularization, nor does it make the resulting training gradient an unbiased failure-risk gradient.

\paragraph{Composed policy and gate.}\label{sec:compose} To deploy the trained specialist with a nominal fallback, we return $\Pi=(\pi_{\rm nom},\pi_{\rm crit},C_\phi,\tau)$: the frozen nominal policy, trained specialist, criticality model, and gate threshold. Let $g_\tau(z)\in\{0,1\}$ indicate that the available context $z$ lies on the high-risk side of the threshold. The action policy is
\begin{equation}
\Pi(a\mid o,z)=g_\tau(z)\pi_{\rm crit}(a\mid o)+[1-g_\tau(z)]\pi_{\rm nom}(a\mid o).
\label{eq:gate}
\end{equation}
This gate routes to one component without observing the rollout outcome. It preserves nominal behavior only on contexts assigned to the nominal policy. Thresholds are selected on validation rollouts and fixed for testing (Table~\ref{tab:random_gate_thresholds}); routing rules appear in Appendix~\ref{app:thresholds}.

Go2 gates on the maximum candidate risk score at the current observation; ManiSkill's selected $\tau=0$ uses the specialist throughout. The imitation-learning domains gate once from the initial scene. LIBERO and RoboTwin require 92-D and 437-D simulator state in addition to the policy's visual inputs; the physical robot uses eight coordinates extracted from RGB. Both policy components remain available during inference.

\section{Experiments}
\label{sec:experiments}

Table~\ref{tab:ablation} compares the frozen nominal policy (Baseline), the policy fine-tuned on randomly collected data (Random), and the gated variants (Random + gate and Ours). Random + gate uses the same nominal policy, criticality model, and routing rule as Ours, with a separately validation-selected threshold. The imitation-learning domains match fine-tuning settings and data quantities. For RL, we distinguish the best observed rates within a common round budget from our final configurations. Round count alone does not match total data or compute; Go2 also uses different hyperparameters. These comparisons therefore evaluate complete training procedures.

\begin{table}[ht]
\centering
\small
\setlength{\tabcolsep}{1.5mm}
\begin{tabular}{@{}lccccc@{}}
\toprule
& & & & \multicolumn{2}{c}{\textbf{Ours} $\downarrow$} \\
\cmidrule(l){5-6}
\textbf{Domain / unit} & \textbf{Baseline} $\downarrow$ & \textbf{Random} $\downarrow$ & \textbf{Random + gate} $\downarrow$ & \textbf{Shared rounds} & \textbf{Final} \\
\midrule
Go2 ($\times10^{-4}$) & 3.58 & 3.57 & 3.38 & 2.48 & \textbf{1.47} \\
ManiSkill (\%) & 5.84 & 3.27 & 3.27 & 2.34 & \textbf{2.08} \\
LIBERO (\%) & 1.87 & 1.82 & 1.68 & \multicolumn{2}{c}{\textbf{1.40}} \\
RoboTwin (\%) & 14.40 & 20.70 & 14.00 & \multicolumn{2}{c}{\textbf{13.30}} \\
Real robot (\%) & 11.15 & 18.05 & 8.53 & \multicolumn{2}{c}{\textbf{3.72}} \\
\bottomrule
\end{tabular}
\caption{Failure rates. For RL, Random and Ours (Shared rounds) report the lowest observed rates within the first 3 Go2 or 12 ManiSkill rounds; Ours (Final) uses Round~6 or 17. Random + gate reports its validation-selected configuration. The other domains use one update. Physical values are weighted estimates from 50 trials.}
\label{tab:ablation}
\end{table}

\begin{table}[ht]
\centering
\small
\setlength{\tabcolsep}{4mm}
\begin{tabular}{@{}lcc@{}}
\toprule
\textbf{Domain} & \textbf{Random + gate} $\boldsymbol{\tau}$ & \textbf{Ours} $\boldsymbol{\tau}$ \\
\midrule
Go2 & 0.35 & 0.50 \\
ManiSkill & 0 & 0 \\
LIBERO (10 / Goal / Object / Spatial) & 0.90 / 0 / 0.35 / 0 & 0.90 / 0 / 0.50 / 0.60 \\
RoboTwin & 0.60 & 0.65 \\
Real robot & 0.15 & 0.15 \\
\bottomrule
\end{tabular}
\caption{Validation-selected gate thresholds, fixed during testing. Go2 entries correspond to the best observed checkpoints: Round~1 for Random + gate and Round~6 for Ours. LIBERO thresholds follow the listed suite order.}
\label{tab:random_gate_thresholds}
\end{table}

\paragraph{Training gains and the role of gating.} Using Ours' thresholds for both arms, Ours lowers failure point estimates relative to Random + gate by 0.32, 0.80, and 4.81 percentage points on LIBERO, RoboTwin, and the physical task (Appendix~\ref{app:random_gate}). These comparisons support the value of criticality-guided training with routing held fixed. Gating also contributes: LIBERO and RoboTwin specialists alone have 2.13\% and 16.30\% failure, versus 1.87\% and 14.40\% for their nominal policies; composition yields 1.40\% and 13.30\%. Thus training and routing play complementary roles. Guided training combines collection and resampling; Appendix~\ref{app:random_gate} discusses evaluation uncertainty and untested variation across training seeds and calibration splits.

\begin{figure}[t]
\centering
\includegraphics[width=0.72\textwidth]{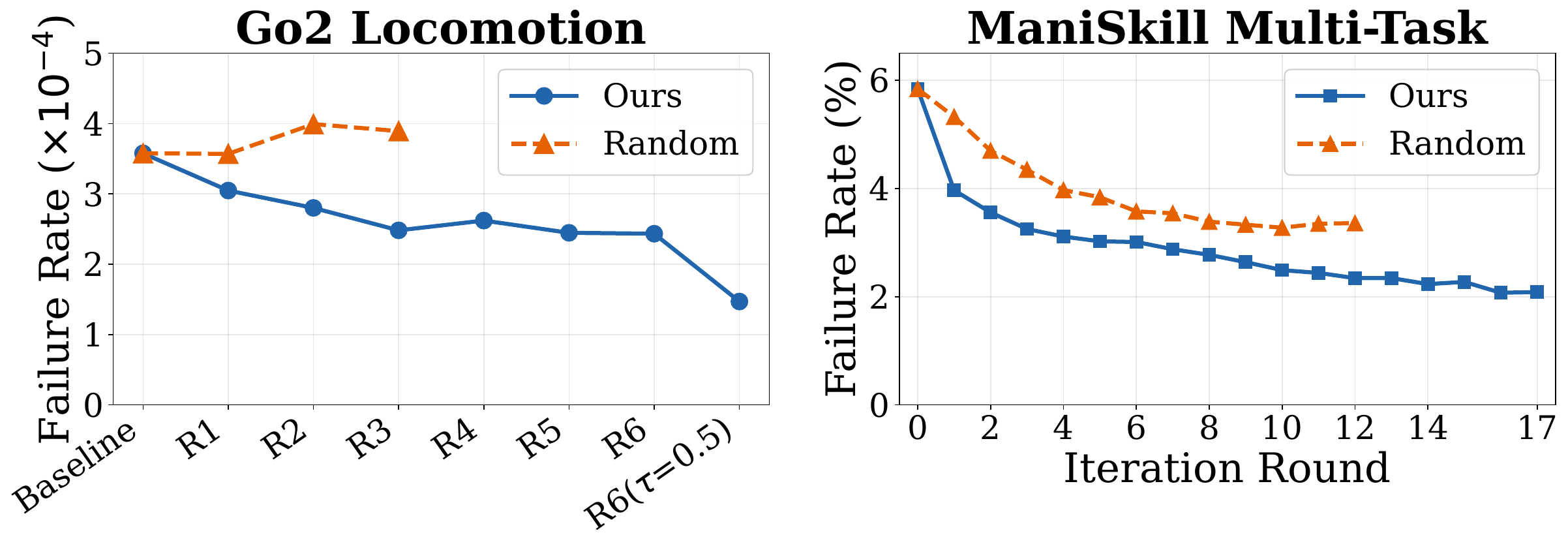}
\caption{Go2 (left) and ManiSkill (right) multi-round failure trends. Blue denotes the composed policy (Ours) and orange fine-tuning on randomly collected data (Random). Only mean curves are shown; intervals are in Appendix~\ref{app:rounds}. The arms end at different round budgets. Go2 rounds 1--6 use $\tau=0.8$. The final point evaluates the same Round~6 specialist with $\tau=0.5$, selected after training on separate validation rollouts. ManiSkill uses $\tau=0$ throughout. Full values are in Appendix~\ref{app:rounds}.}
\label{fig:trend}
\end{figure}

\paragraph{Go2 locomotion.} A PPO-pretrained 48-D state policy controls 12 joint-target corrections in MuJoCo \citep{todorov2012mujoco}. Within the shared three-round budget, Ours reaches $2.48\times10^{-4}$ at Round~3, versus Random's minimum of $3.57\times10^{-4}$ at Round~1. At exactly Round~3, Random is $3.89\times10^{-4}$. Continuing Ours to Round~6 gives $2.43\times10^{-4}$ at $\tau=0.8$; validation-selected $\tau=0.5$ then gives $1.47\times10^{-4}$ without another policy update. Gating the Round~1 Random checkpoint gives $3.38\times10^{-4}$ at validation-selected $\tau=0.35$.

With $\tau$ fixed at 0.8, guided updates lower failure from the baseline's $3.58\times10^{-4}$ to $2.43\times10^{-4}$, with most of this decrease reached by Round~3. Subsequent gate calibration adds a substantial reduction to $1.47\times10^{-4}$. Each round combines new and historical data from a $10^4$-cell terrain grid. The arms differ in epochs, loss coefficients, and accumulated data (Appendix~\ref{app:go2}), so equal round counts do not isolate collection.

\paragraph{ManiSkill manipulation.} A shared mixture-of-experts (MoE) policy is trained across four tasks. We report StackCube and PegInsertionSide under forces from an $11^3$ grid \citep{tao2025maniskill3}. At the same Round~12, average failure is 2.34\% for Ours versus 3.36\% for Random. Both use $\tau=0$, so the specialist acts throughout and the improvement comes from training without selective gating. Within the first 12 rounds, the lowest observed rates are 2.34\% for Ours (Round~12) and 3.27\% for Random (Round~10). Ours finishes at 2.08\% in Round~17, with an overall observed minimum of 2.07\% in Round~16.

Both arms use the same per-round optimizer settings and 25,000 evaluation rollouts per task. This comparison demonstrates a training gain before the five additional guided updates. Equal rounds do not establish equal end-to-end cost, and this comparison does not isolate predictor refresh. Figure~\ref{fig:trend} and Appendix~\ref{app:rounds} retain the complete curves.

\paragraph{LIBERO.} We adapt the imported SimVLA checkpoint \citep{luo2026simvla} for 40K steps on the full demonstrations to resolve weight-map incompatibilities. Both continuations start there and use 40K BC iterations on about 1,000 demonstration slots selected from the 2,000-example official training pool. The criticality model is trained on 40,000 separate self-generated scenes. Across four suites, Ours has 1.40\% failure and Random + gate 1.68\%. The mean difference is 0.28 percentage points; its paired interval includes zero (Appendix~\ref{app:random_gate}). On Spatial, however, Ours reaches 1.73\%, below the 2.13\% nominal rate but above Random + gate's 1.60\% (Table~\ref{tab:libero}).

Each policy executes the same 2,000 official task starts three times. The predictor scenes and BC demonstrations are disjoint from evaluation starts. Suite results and component diagnostics are in Appendix~\ref{app:domain_results} and Appendix~\ref{app:libero}.

\paragraph{RoboTwin.} From the StarVLA Qwen3-VL-OFT RoboTwin2-All checkpoint \citep{starvla2026}, both variants receive matched 40K-step action-head BC with 10\% full replay. The guided proposal selects 5,000 slots representing 2,574 unique episodes across 50 tasks. On the shared 1,000-episode manifest, failure is 13.30\% for Ours and 14.00\% for Random + gate. Against Baseline, Ours repairs 19 failures and introduces eight regressions (net 11). Paired statistical analysis is given in Appendix~\ref{app:random_gate}.

To compare the components under identical routing, we also fix both thresholds at 0.65. This differs from Random + gate's selected $\tau=0.60$ and gives 14.10\% for Random + gate: on the same 124 routed episodes, Random repairs 16 failures and introduces 13, versus 19 and eight for the guided specialist. Appendix~\ref{app:robotwin} describes the initial-state interface.

\begin{figure}[ht]
\centering
\includegraphics[width=\textwidth]{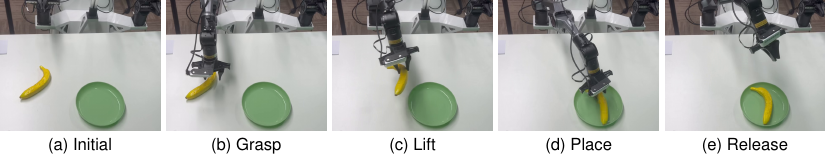}
\caption{Physical banana pick-and-place on the AgileX Piper robot. Frames (a)--(e) show one test rollout in chronological order.}
\label{fig:real_robot_sequence}
\end{figure}

\paragraph{Physical manipulation.} On AgileX Piper, a $\pi_{0.5}$ policy \citep{physicalintelligence2025pi05} performs banana pick-and-place (Figure~\ref{fig:real_robot_sequence}). Both continuations use 100 expert demonstrations and 10,000 updates from the same checkpoint. At the same validation-selected $\tau=0.15$, weighted failure estimates are 3.72\% for Ours and 8.53\% for Random + gate. Their mean difference is 4.81 percentage points. These estimates remain imprecise: the 50 trials have weight effective sample size 12.44 (Appendix~\ref{app:random_gate}).

The criticality model trains on 99 labeled episodes using initial-image object coordinates. It scores 10,000 candidate scenarios for expert-data collection; failed policy rollouts provide predictor labels, not BC supervision. In point estimates, the specialist alone lowers weighted failure from 11.15\% to 6.05\%, realizing most of the total reduction: 5.10 of 7.43 percentage points before gating, with a further 2.33 points from composition (Appendix~\ref{app:real}).

\subsection{Evaluation Protocol}

Go2 and the physical task use unnormalized importance-weighted evaluation. Other domains use direct failure frequencies. LIBERO and RoboTwin use shared evaluation manifests, and physical comparisons use paired scenes. Best-round values are descriptive minima of the recorded test curves, not checkpoints selected on independent validation data, and may be optimistic due to selection. Gate thresholds, by contrast, are selected on validation rollouts and fixed for testing. Paired uncertainty and clustering checks are in Appendix~\ref{app:random_gate}.

\paragraph{Acquisition cost and deployment inputs.} The method requires outcome-labeled rollouts, suitable policy-training data, and a risk input available at inference. LIBERO and RoboTwin use simulator state; the physical system uses task-specific RGB coordinates. Matched BC budgets exclude predictor collection and calibration, including LIBERO's 40,000 predictor scenes. Reported predictor timings exclude input extraction, routing, and dual-policy memory. Appendix~\ref{app:cost_ledger} records known budgets and unmeasured costs.

\section{Conclusion}

We introduced a policy-level recursive self-improvement framework centered on training guided by a criticality world model, extending across RL-pretrained locomotion and manipulation and BC-trained VLA policies. By directing data collection toward remaining failures and resampling for domain-specific updates, the framework trains specialists that a risk-based gate combines with a nominal fallback. Empirically, ManiSkill improves without selective gating, and the physical specialist realizes most of the observed reduction before gating. Fixed-gate comparisons further favor guided training, while calibration adds a substantial gain in Go2. Together, these findings support criticality-guided training as the core improvement mechanism, complemented by policy composition. Separate effects of resampling and predictor refresh, and reproducibility across VLA training runs, remain unresolved.


\clearpage
\subsection*{AI Use Statement}

In this work, we used generative AI tools for manuscript editing and condensation, checking
and presenting mathematical derivations, and implementation and debugging of research code. We have reviewed all AI-assisted work. We take responsibility for the final content of this work, including text, claims or artifacts produced with the aid of generative AI.

\subsection*{Reproducibility Statement}

Section~\ref{sec:theory} gives the sampling assumptions and variance result, with proofs in Appendix~\ref{app:theory_proofs}. Sections~\ref{sec:method} and~\ref{sec:experiments} describe the proposal,
specialist update, policy composition, and evaluations. The appendix contains per-domain
tables (Appendix~\ref{app:domain_results}), the unnormalized evaluation
estimator and weighted uncertainty procedure (Appendix~\ref{app:is_eval}),
perturbations and training protocols (Appendices~\ref{app:go2}--\ref{app:real}),
predictor architectures (Appendix~\ref{app:criticality_summary}), and complete
multi-round results (Appendix~\ref{app:rounds}). Simulation results use at least
1,000 evaluation episodes per model. The physical evaluation uses 50 weighted
trials. Intervals are reported where available.

\subsection*{Ethics Statement}

The physical experiment uses a lightweight pick-and-place platform in a fixed
workspace, without human subjects or personal data. Failure-focused rollout collection
can increase collisions or dropped objects. We bounded the
perturbations and operated away from people. Component selection can also fail
because the score is learned from finite rollout data. Its threshold therefore
requires separate validation for safety-critical deployment. These empirical
results are not a safety certification for uncontrolled or human-populated
environments.

\bibliographystyle{iclr2027_conference}
\bibliography{references}

\clearpage
\appendix
\section{Additional Domain Results}
\label{app:domain_results}
\suppressfloats[t]

Throughout these tables, Specialist only denotes the criticality-guided specialist evaluated without the gate; Ours denotes its composition with the frozen nominal policy.

\begin{table}[!htbp]
\centering
\footnotesize
\setlength{\tabcolsep}{0.7mm}
\begin{tabular}{@{}lcccccc@{}}
\toprule
\textbf{Model} & \textbf{10} $\downarrow$ & \textbf{Goal} $\downarrow$ & \textbf{Object} $\downarrow$ & \textbf{Spatial} $\downarrow$ & \textbf{Avg.} $\downarrow$ & \textbf{vs. baseline} $\downarrow$ \\
\midrule
Baseline & 2.93 & 1.87 & 0.53 & 2.13 & 1.87 & --- \\
Random & 3.27 & 1.80 & 0.60 & 1.60 & 1.82 & $-2.7\%$ \\
Random + gate & 2.93 & 1.80 & 0.40 & 1.60 & 1.68 & $-9.8\%$ \\
Specialist only & 4.20 & 1.40 & 1.00 & 1.93 & 2.13 & $+14.3\%$ \\
\textbf{Ours} & \textbf{2.20} & \textbf{1.40} & \textbf{0.27} & 1.73 & \textbf{1.40} & $-25.0\%$ \\
\bottomrule
\end{tabular}
\caption{LIBERO failure rate (\%) by suite, averaged over three evaluation
runs per suite (6,000 episodes per model). Random denotes fine-tuning on randomly collected data; Random + gate uses its validation-selected suite thresholds (Appendix~\ref{app:random_gate}).}
\label{tab:libero}
\end{table}

\begin{table}[ht]
\centering
\begin{tabular}{@{}cccccc@{}}
\toprule
\textbf{Domain} & \textbf{Policy} & \textbf{Perturb.} & \textbf{Obs.} & \textbf{Obs.\ Dim.} & \textbf{Act.\ Dim.} \\
\midrule
Go2 Locomotion & MLP (0.19M) & Terrain & State & 48 & 12 \\
ManiSkill (Stack) & Mixture-of-Experts (0.32M) & Force 3D & State & 48 & 8 \\
ManiSkill (Peg) & Mixture-of-Experts (0.32M) & Force 3D & State & 43 & 8 \\
LIBERO & SimVLA & Init.\ state & RGB & 3$\times$384$\times$384 & 7 \\
RoboTwin & StarVLA & Init.\ state & RGB & 3$\times$240$\times$320 & 14 \\
Real (Banana) & $\pi_{0.5}$ & Obj.\ pose & RGB & 3$\times$960$\times$540 & 7 \\
\bottomrule
\end{tabular}
\caption{Overview of experimental domains. ``Obs.'' describes the action policy. LIBERO and RoboTwin additionally expose initial simulator scene state to the gate.}
\label{tab:domains}
\end{table}

\begin{table}[ht]
\centering
\small
\setlength{\tabcolsep}{2.4mm}
\begin{tabular}{@{}lcccc@{}}
\toprule
\textbf{Domain} & \textbf{Policy} & \textbf{Selection variable} & \textbf{Gate input} & \textbf{Update rounds} \\
\midrule
Go2 & State MLP & Terrain / step & State & 6 \\
ManiSkill & Multi-task MoE & Force / step & State & 17 \\
LIBERO & SimVLA & Demo / scene & Simulator state & 1 \\
RoboTwin & StarVLA & Demo / scene & Simulator state & 1 \\
Real robot & $\pi_{0.5}$ & Object pose & RGB coordinates & 1 \\
\bottomrule
\end{tabular}
\caption{Evaluation interfaces. The LIBERO and RoboTwin gates receive initial simulator state in addition to the VLA policy observations.}
\label{tab:domains_main}
\end{table}

\begin{table}[ht]
\centering
\footnotesize
\setlength{\tabcolsep}{1.0mm}
\begin{tabular}{@{}lcc@{}}
\toprule
\textbf{Model} & \textbf{Failure Rate} $\downarrow$ & \textbf{vs.\ Baseline} $\downarrow$ \\
\midrule
Baseline & $3.58 \times 10^{-4}$ & --- \\
\textbf{Ours} & $\mathbf{1.47 \times 10^{-4}}$ & \textbf{$-$58.9\%} \\
\bottomrule
\end{tabular}
\caption{Go2 locomotion results. Evaluated using 119,990 baseline and 120,000 composed-policy importance-sampled rollouts.}
\label{tab:go2}
\end{table}

\begin{table}[ht]
\centering
\footnotesize
\setlength{\tabcolsep}{1.2mm}
\begin{tabular}{@{}lcccc@{}}
\toprule
\textbf{Model} & \textbf{Stack} $\downarrow$ & \textbf{Peg} $\downarrow$ & \textbf{Avg.} $\downarrow$ & \textbf{vs.\ Baseline} $\downarrow$ \\
\midrule
Baseline & 5.01 & 6.67 & 5.84 & --- \\
Random (Round 10, best) & 1.88 & 4.66 & 3.27 & $-$43.9\% \\
Random (Round 12, final) & 1.94 & 4.77 & 3.36 & $-$42.5\% \\
Ours (Round 12) & 1.77 & 2.91 & 2.34 & $-$59.9\% \\
Ours (Round 16, best) & 1.48 & 2.66 & 2.07 & $-$64.5\% \\
\textbf{Ours (Round 17, final)} & \textbf{1.48} & \textbf{2.68} & \textbf{2.08} & \textbf{$-$64.3\%} \\
\bottomrule
\end{tabular}
\caption{ManiSkill multi-task manipulation results. Failure rate (\%) for StackCube (Stack) and PegInsertionSide (Peg). Each updated configuration uses 25,000 rollouts per task. Round~12 compares equal update counts; best values are descriptive minima from the recorded test curves.}
\label{tab:maniskill}
\end{table}

\begin{table}[ht]
\centering
\footnotesize
\setlength{\tabcolsep}{1.2mm}
\begin{tabular}{@{}lcc@{}}
\toprule
\textbf{Model} & \textbf{Failure (\%)} $\downarrow$ & \textbf{vs.\ Baseline} $\downarrow$ \\
\midrule
Baseline              & 14.40 & --- \\
Random                & 20.70 & +43.8\% \\
Random + gate         & 14.00 & $-$2.8\% \\
Specialist only       & 16.30 & +13.2\% \\
\textbf{Ours} & \textbf{13.30} & \textbf{$-$7.6\%} \\
\bottomrule
\end{tabular}
\caption{RoboTwin failure rate (\%). Strict common evaluation over 50 tasks $\times$ 20 episodes = 1000 episodes.}
\label{tab:robotwin}
\end{table}

\begin{table}[ht]
\centering
\footnotesize
\setlength{\tabcolsep}{1.2mm}
\begin{tabular}{@{}lcc@{}}
\toprule
\textbf{Model} & \textbf{Failure (\%, 95\% CI)} $\downarrow$ & \textbf{vs.\ Baseline} $\downarrow$ \\
\midrule
Baseline              & 11.15 [6.22, 16.72] & --- \\
Random                & 18.05 [9.43, 28.25] & +61.9\% \\
Random + gate         & 8.53 [4.59, 13.12] & $-$23.5\% \\
Specialist only       & 6.05 [1.83, 11.93] & $-$45.7\% \\
\textbf{Ours} & \textbf{3.72 [1.31, 6.63]} & \textbf{$-$66.6\%} \\
\bottomrule
\end{tabular}
\caption{Real-world banana-on-plate failure rates from 50 importance-sampled test trials. Brackets give 95\% percentile intervals from paired episode bootstrap. Relative changes use point estimates.}
\label{tab:real}
\end{table}

\FloatBarrier
\section{Information Density in Criticality-Guided Sampling}
\label{app:toy}

Our framework selects more failure-prone scenarios and then resamples the selected buffer by inverse-proposal weights. The exact identity in Equation~\ref{eq:unbiased} applies before finite-buffer normalization, whereas the realized training buffer only approximates that target. The benefit sought is broader coverage of informative scenarios that uniform collection may miss under a fixed budget.

Repeated exposure to one critical scenario cannot cover failure patterns absent from it. A more varied selected pool can improve critical coverage while reducing nominal coverage under a fixed budget. Proposition~\ref{prop:rho} describes the variance of an idealized average under a specified average pairwise gradient correlation $\bar\rho$. We have not measured this correlation in the toy study or robot experiments. The toy study instead illustrates the coverage trade-off.

\subsection{Sampling Identities and Variance Proof}
\label{app:theory_proofs}

For fixed $\theta$ and a fixed finite candidate pool, assume $g_\theta(s)$ has finite second moment. Cauchy--Schwarz gives
\begin{equation}
\left(\sum_s P(s)\|g_\theta(s)\|_2\right)^2
\leq \left(\sum_s \frac{P(s)^2\|g_\theta(s)\|_2^2}{q(s)}\right)
\left(\sum_s q(s)\right).
\end{equation}
Equality holds when $q(s)\propto P(s)\|g_\theta(s)\|_2$, proving Equation~\ref{eq:qstar} on the nonzero-gradient support. A positive exploration floor preserves support but generally changes this optimum. For stochastic gradients $\widehat g$ conditional on $s$, replace $\|g_\theta(s)\|_2$ with $\sqrt{\mathbb E[\|\widehat g\|_2^2\mid s]}$.

More explicitly, let $\mathcal C=\{s_1,\ldots,s_N\}$ contain independent draws from $P_0$, with $p(i\mid\mathcal C)=1/N$. Conditional on the pool, draw indices $I_1,\ldots,I_m$ independently from $q(\cdot\mid\mathcal C)$. For any fixed $f$ independent of pool construction,
\begin{equation}
\mathbb E_{\mathcal C,I_{1:m}}\left[\frac1m\sum_{k=1}^{m}\frac{p(I_k\mid\mathcal C)}{q(I_k\mid\mathcal C)}f(s_{I_k})\right]=\mathbb E_{s\sim P_0}[f(s)].
\label{eq:two_stage_is}
\end{equation}
This result requires the stated independent draws, positive support, and actual $p/q$ correction. Sampling without replacement requires inclusion probabilities instead. It does not make a finite, self-normalized behavior-cloning buffer unbiased.

For Equation~\ref{eq:two_stage_is}, conditioning on $\mathcal C$ reduces each weighted summand's expectation to $N^{-1}\sum_{i=1}^N f(s_i)$. Taking expectation over the independent draws $s_i\sim P_0$ gives $\mathbb E_{P_0}f$. This reasoning fixes the integrand before sampling. It does not assert unbiasedness for parameters fitted to the same realized buffer.

By contrast, the conditional expectation of a draw resampled from a selected buffer with normalized weights is $\sum_k w_k f(s_{I_k})/\sum_k w_k$. This is a ratio of random sums. Its expectation need not equal $\mathbb E_P f$ at finite buffer size. For a fixed finite pool, independent proposal draws and positive support make it converge to the pool expectation as the selected buffer grows.

\begin{proof}[Proof of Proposition~\ref{prop:rho}]
Pool~A has mean $\bar Z_A=Z$, so $\operatorname{Var}(\bar Z_A)=\sigma^2$. For Pool~B, define $\bar\rho=2\sum_{i<j}\operatorname{Corr}(Z_i,Z_j)/[N(N-1)]$. Expanding the covariance sum gives
\begin{equation}
\operatorname{Var}(\bar Z_B)=\frac{N\sigma^2+2\sum_{i<j}\operatorname{Cov}(Z_i,Z_j)}{N^2}
=\frac{\sigma^2}{N}\bigl[1+(N-1)\bar\rho\bigr].
\end{equation}
Dividing by $\sigma^2$ yields the stated ratio. The admissible range is $-1/(N-1)\leq\bar\rho\leq1$ for $N\geq2$.
\end{proof}

\subsection{Toy Example Design and Hypothesis}

\begin{table}[t]
\centering
\small
\setlength{\tabcolsep}{1.5mm}
\begin{tabular}{@{}lcccc@{}}
\toprule
\textbf{Pool} & \textbf{Critical Scenarios} & \textbf{Draws per Scenario} & \textbf{Nominal Scenarios} & \textbf{Draws per Scenario} \\
\midrule
A & 1 distinct  & 10 & 10 distinct & 10  \\
B & 10 distinct & 1  & 1 distinct  & 100 \\
\bottomrule
\end{tabular}
\caption{Controlled training pools. Both have 110 draws per batch and a 10:100
critical-to-nominal ratio, but different scenario coverage.}
\label{tab:pool_construction}
\end{table}

Consider the two training pools of Table~\ref{tab:pool_construction}, which share an identical batch size and critical-to-nominal ratio.

\begin{itemize}
\item \textbf{Pool A} contains 1 critical scenario and 10 nominal scenarios. Each is sampled 10 times per epoch (batch size = $1 \times 10 + 10 \times 10 = 110$).
\item \textbf{Pool B} contains 10 distinct critical scenarios and 1 nominal scenario. Each critical scenario is sampled once per epoch. The single nominal scenario is sampled 100 times (batch size = $10 \times 1 + 1 \times 100 = 110$).
\end{itemize}

Both pools produce 10 critical and 100 nominal samples per batch, but they define different empirical training distributions. Pool~B covers more critical scenarios and fewer nominal scenarios. Equal class proportions do not make either pool an unbiased estimate of the other's objective. We therefore expect Pool~B to learn more of the ten critical patterns and Pool~A to retain better nominal coverage. This trade-off motivates retaining a nominal policy alongside the specialist, although the experiment does not measure gradient correlations.

\subsection{Experimental Validation}

To examine the coverage trade-off, we construct a controlled 2-D binary classification experiment. Nominal Gaussian-blob scenarios have labels determined by $\text{sign}(x)$. Critical ``exception'' scenarios lie near the decision boundary $x \approx 0$ and have alternating labels that violate the nominal pattern. A \emph{scenario} is a distribution, not one fixed point. ``Sampling $N$ times'' means $N$ independent draws from that distribution.

\subsubsection{Experimental Design}

Both pools train an MLP with two hidden layers (2$\to$128$\to$128$\to$1, batch normalization (BatchNorm), ReLU) with binary cross-entropy (BCE) loss for 500 epochs using full-batch AdamW (lr~$=$~$3\times10^{-3}$, weight decay~$=$~$10^{-5}$). Results summarize 10 random seeds.

The primary metric is accuracy on the nine critical regions absent from Pool~A's training pool. Because Pool~B includes these regions during training, the comparison measures the effect of covering more of the evaluated critical scenarios under the fixed sample budget, rather than generalization by both pools to unseen modes.

\subsubsection{Results}

\begin{figure}[ht]
\centering
\includegraphics[width=0.78\textwidth]{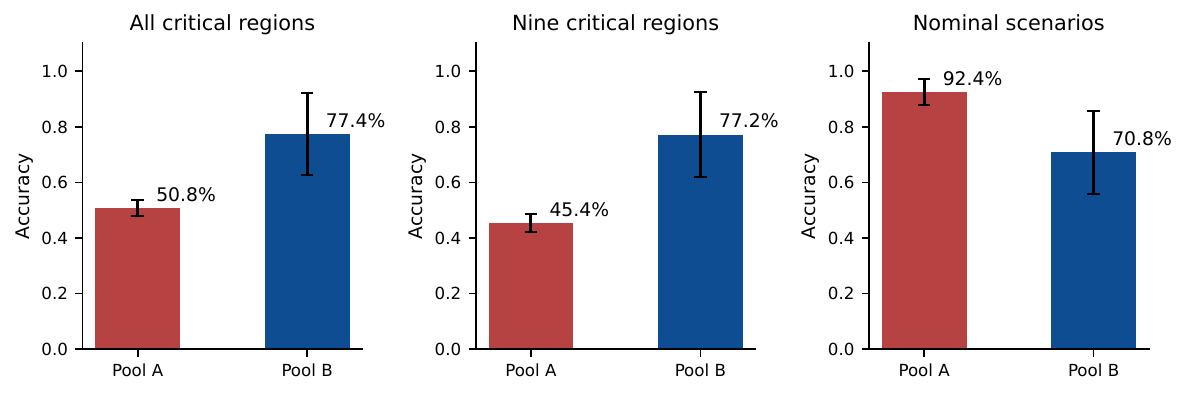}
\caption{Controlled coverage results with error bars (mean $\pm$ one standard deviation across ten seeds). The main-text version shows means only.}
\label{fig:toy_with_sd}
\end{figure}

Figure~\ref{fig:main_results} and Table~\ref{tab:main_results} report 77.2\% accuracy for Pool~B on those nine critical regions, versus 45.4\% for Pool~A. The 31.8 percentage-point difference occurs with the same class ratio and batch size (standardized mean difference $d=2.89$, using the pooled marginal standard deviation across ten seeds).

\begin{table}[ht]
\centering
\small
\setlength{\tabcolsep}{1.5mm}
\begin{tabular}{lcc}
\toprule
\textbf{Metric} & \textbf{Pool A} & \textbf{Pool B} \\
\midrule
Overall Critical Accuracy & 50.79\% $\pm$ 3.02\% & \textbf{77.36\%} $\pm$ 14.73\% \\
Nine-Region Critical Accuracy\textsuperscript{$\star$} & 45.38\% $\pm$ 3.30\% & \textbf{77.18\%} $\pm$ 15.18\% \\
Nominal Accuracy & \textbf{92.41\%} $\pm$ 4.67\% & 70.79\% $\pm$ 14.93\% \\
\bottomrule
\end{tabular}
\caption{Controlled classification results (mean $\pm$ std across 10 seeds). \textsuperscript{$\star$}The primary metric covers the nine critical regions absent from Pool~A but included in Pool~B. The primary metric has standardized mean difference $d=2.89$.}
\label{tab:main_results}
\end{table}

\begin{figure}[ht]
\centering
\includegraphics[width=0.77\textwidth]{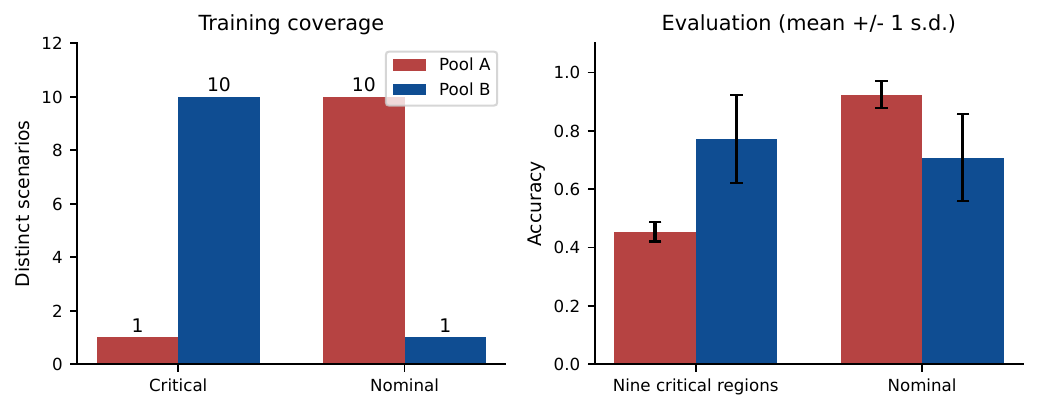}
\caption{Distinct-scenario coverage in the controlled task. Pool~B has higher critical accuracy by 31.8 percentage points, while Pool~A has higher nominal accuracy by 21.6 points. Batch size and class ratio are fixed, but the pools cover different numbers of scenarios in each category.}
\label{fig:density_symmetry}
\end{figure}

This critical-accuracy gain comes with lower nominal accuracy (Figure~\ref{fig:density_symmetry}). Pool~A has 10 distinct nominal scenarios versus 1 in Pool~B and reaches 92.4\% rather than 70.8\% nominal accuracy. Pool~B has 10 distinct critical scenarios versus 1 in Pool~A and reaches 77.2\% rather than 45.4\% critical accuracy. The class ratio and batch size are held fixed while the scenario sets differ.

\subsubsection{Learning Dynamics}

\begin{figure}[ht]
\centering
\includegraphics[width=0.59\textwidth]{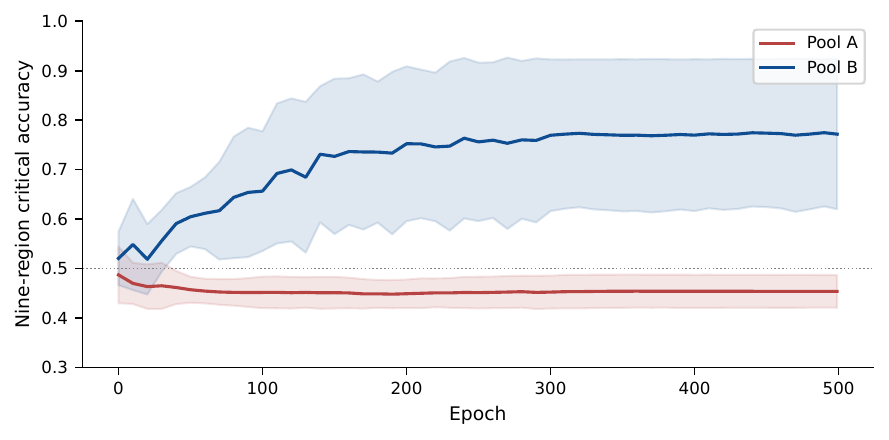}
\caption{Learning dynamics on the nine regions absent from Pool~A (mean $\pm$ 1 s.d., 10 seeds). After initially learning the nominal pattern, Pool~A remains near chance on these regions while Pool~B improves.}
\label{fig:learning_dynamics}
\end{figure}

Figure~\ref{fig:learning_dynamics} tracks accuracy on the nine regions absent from Pool~A over training. Both pools initially learn the dominant nominal pattern (epochs 0--20), and accuracy on these regions falls as the model approaches the simple decision boundary $x \approx 0$. Their trajectories then diverge.

\begin{itemize}
\item \textbf{Pool~A stagnates} during epochs 20--500. Repeated draws from its one critical region do not expose it to the other nine. Accuracy there stays near chance even as training loss approaches zero.
\item \textbf{Pool~B improves} over the same epochs. Its pool includes all ten critical regions, and accuracy on the nine-region metric rises from $\sim$50\% to 77.2\%.
\end{itemize}

\subsubsection{Summary}

With the same 10:100 class ratio and 110-example batch size, covering ten critical scenarios yields higher accuracy on the nine regions absent from Pool~A than repeatedly sampling one region (standardized mean difference $d=2.89$ across ten seeds). However, Pool~B sacrifices nominal coverage. The result therefore demonstrates a coverage trade-off in this task without measuring the gradient correlations in Proposition~\ref{prop:rho}.

\section{Importance-Sampled Evaluation}
\label{app:is_eval}

\subsection{Motivation}

For policies with very low failure rates, naive Monte Carlo evaluation under uniform sampling is prohibitively expensive. At Go2's baseline failure rate of $\mu \approx 3.6 \times 10^{-4}$, roughly 2,800 rollouts yield one failure in expectation, and about 280,000 yield 100. Precise estimates therefore require large rollout budgets. Physical trials are costly as well. A budget of 50 real-robot episodes would yield only a handful of failures under uniform sampling, producing noisy estimates.

\subsection{Estimator}

To use the evaluation budget more efficiently, we use the proposal in Equations~\ref{eq:q_rl}--\ref{eq:q_il}, whose $\varepsilon$ mixture ensures positive support. For $N$ episodes with scenarios $s_i\sim q$, the unnormalized importance sampling (IS) estimator is

\begin{equation}
\hat{\mu}_{\text{IS}} = \frac{1}{N}\sum_{i=1}^{N} W_i \cdot \mathbf{1}_{\text{f}}(s_i),
\label{eq:is_estimator}
\end{equation}

Here $W_i=\prod_t w(s_{i,t})$ is the product of per-step weights for a multi-step rollout. For initial-state evaluation, $W_i=w(s_0^{(i)})$.

\subsection{Unbiasedness}

Under the support condition $q(s) > 0$ wherever $P(s) > 0$,

\begin{equation}
\begin{aligned}
\mathbb{E}_{s \sim q}\bigl[w(s) \cdot \mathbf{1}_{\text{f}}(s)\bigr]
&= \int q(s) \frac{P(s)}{q(s)} \mathbf{1}_{\text{f}}(s) \, ds \\
&= \int P(s) \mathbf{1}_{\text{f}}(s) \, ds = \mu.
\end{aligned}
\label{eq:is_unbiased}
\end{equation}

Within each domain, the importance-weighted and uniform-sampling estimators target the same candidate-pool failure probability $\mu$ when their respective sampling probabilities are correct. Although the unnormalized estimator is unbiased for any positive-support $q$, its variance depends on the proposal. A finite empirical candidate pool need not equal the full physical deployment distribution.

\subsection{Variance Reduction}

For deterministic outcomes conditional on $s$, the zero-variance proposal would concentrate according to $P(s)\mathbf{1}_{\rm f}(s)$, if the failures were already known. With stochastic outcomes its variance-minimizing form depends on the conditional second moment. $C_\phi$ is an empirical risk-ranking proxy, not either oracle proposal. Thus high-risk sampling can improve precision, but the amount of variance reduction must be assessed from realized weights and outcomes rather than assumed from the proposal form.

\subsection{Implementation}

\begin{enumerate}
\item Fix the trained $C_\phi$, policy components, and validation-selected gate before evaluation.
\item Sample initial scenes or per-step perturbations from the $\varepsilon$-mixture proposal $q$, retaining the actual sampling probabilities.
\item Execute $N$ episodes, recording binary outcomes and the corresponding proposal ratios.
\item Compute $\hat{\mu}_{\text{IS}}$ via Eq.~\ref{eq:is_estimator}. For Go2, 1500 parallel workers schedule 80 episodes each. Completed configurations contain 119,920--120,000 importance-sampled episodes, with 95\% bootstrap confidence intervals. For the real robot, $N = 50$ importance-sampled episodes with per-episode importance weights.
\end{enumerate}

For a reported weighted estimate, uncertainty is calculated from the
individual weighted outcomes $Z_i=W_i\mathbf{1}_{\rm f}(s_i)$, rather than from
the rounded aggregate rate. A direct standard-error estimate is
$\sqrt{\sum_i(Z_i-\bar Z)^2/[N(N-1)]}$. Our real-robot intervals instead use
100,000 percentile-bootstrap resamples of whole episodes, recomputing the
unnormalized estimator $N^{-1}\sum_i Z_i$ each time. Policies tested on the
same scenarios retain their pairing within each resample. The effective
sample size of the 50 recorded real-robot weights is
$N_{\rm eff}=(\sum_i W_i)^2/\sum_i W_i^2=12.44$. These intervals reflect trial
sampling under a fixed proposal and gate threshold. They do not account for
threshold selection or uncertainty in the learned criticality model.

The same criticality model serves training-time selection, the internal gate,
and evaluation, so no separate evaluation-specific model is needed.

\section{Per-Domain Gate Thresholds}
\label{app:thresholds}

For an available context $z$, component selection compares a risk statistic with $\tau$ and chooses $\pi_{\rm crit}$ on the high-risk side, or $\pi_{\rm nom}$ otherwise. Go2 uses the maximum $C_\phi(o_t,u)$ over candidate terrain perturbations $u$ at the current observation $o_t$. The initial-scene domains use $C_\phi(s_0)$. The equality convention varies by domain and is recorded below. LIBERO and RoboTwin require simulator state at reset (92-D padded and 437-D object state, respectively), beyond the action policy's inputs. The physical system uses eight coordinates extracted from its initial red-green-blue (RGB) image. Both policy components remain available at inference, with associated memory and input costs. Routing can preserve nominal behavior only where it selects the appropriate component.

Table~\ref{tab:thresholds} records the thresholds used for the reported composed-policy configurations. In the two multi-round RL experiments, thresholds were preselected for round-wise evaluation and revisited after the final update. Candidate thresholds were compared on validation rollouts with scenarios or random seeds distinct from those used for the reported test results. Go2 used $\tau=0.8$ through Round~6, then selected $\tau=0.5$ for its reported endpoint using the same trained specialist. ManiSkill retained its preselected $\tau=0$. In LIBERO, RoboTwin and the physical task, candidate thresholds were compared on the validation scenes that supplied the criticality training data, with both component policies evaluated on the same scenes; those validation scenes are disjoint from the reported test scenes. These are the best observed settings among those tested, rather than global optima over all possible thresholds.

\begin{table}[ht]
\centering
\small
\setlength{\tabcolsep}{1.5mm}
\begin{tabular}{@{}lcl@{}}
\toprule
\textbf{Domain} & $\boldsymbol{\tau}$ & \textbf{Specialist when} \\
\midrule
Go2               & 0.50 & maximum candidate risk score $> \tau$ \\
ManiSkill         & 0    & All decisions \\
LIBERO-10         & 0.90 & risk score $> \tau$ \\
LIBERO-Goal       & 0    & All decisions \\
LIBERO-Object     & 0.50 & risk score $> \tau$ \\
LIBERO-Spatial    & 0.60 & risk score $> \tau$ \\
RoboTwin          & 0.65 & risk score $\geq \tau$ \\
Real robot        & 0.15 & risk score $\geq \tau$ \\
\bottomrule
\end{tabular}
\caption{Thresholds for the reported composed-policy endpoints. The specialist acts under the listed condition, and the frozen nominal policy acts otherwise.}
\label{tab:thresholds}
\end{table}

At inference, component selection uses the current risk score without access to the episode outcome. Go2 and ManiSkill select their final thresholds on separate validation rollouts, while LIBERO, RoboTwin and the physical task select theirs on the validation scenes used to train the criticality model, where both component policies were evaluated on the same scenes. Those validation scenes are disjoint from the reported test scenes. These procedures separate threshold selection from test evaluation, although stability across multiple calibration splits remains unmeasured.

\section{Go2 Quadruped Locomotion --- Additional Details}
\label{app:go2}

\subsection{Criticality Model Data Construction}

The Go2 criticality model is a three-layer MLP (56$\to$256$\to$256$\to$2) with ReLU activations, taking a 56-dimensional input (36-D robot state, 16-D heightmap, 4-D terrain parameters). Its two-class output supplies a failure-class risk score for sampling and routing. The labeling strategy is as follows.

\begin{itemize}
\item In success episodes, all timesteps are labeled negative (class 0), with steps downsampled at a 1:100 ratio to mitigate class imbalance.
\item In failure episodes, only the final 4 timesteps are labeled positive (class 1), capturing the immediate precursors of failure rather than labeling the entire episode as a failure. Remaining timesteps are labeled negative and downsampled.
\item Failure episodes reaching the 40-step horizon are truncated to the frame where the robot actually stops before applying the last-four-step rule.
\end{itemize}

Terrain collection normalizes candidate risk scores directly, then adds a 0.01 uniform mixture when the maximum score exceeds 0.3 and the cumulative-weight guard permits biased sampling. Otherwise the proposal is uniform. The proposal activation cutoff differs from the component-selection threshold in Table~\ref{tab:thresholds}. Each recorded weight uses the proposal actually applied at that step.

The model is trained for 100 epochs with CrossEntropyLoss (Adam, lr=$10^{-3}$, batch size 512) and selected by validation area under the precision--recall curve (PR-AUC).

\subsection{Specialist Training}

The PPO-pretrained correction policy is updated using Filter-BC. Return filtering selects higher-return examples, and exponential advantage weights emphasize actions with larger estimated advantage. A squared-error BC anchor limits drift. A hinge penalty discourages failure-associated actions, a Huber loss trains the value function, and an action-range penalty limits extrapolation. Normalized trajectory-ratio resampling approximately corrects the collection proposal before these domain-specific losses are applied.

Criticality-guided training and fine-tuning on randomly collected data use different hyperparameters. The latter uses 30 epochs, advantage temperature 0.5, BC coefficient 0.5, failure-penalty coefficient 3.0, and margin 1.0. The guided rounds use 30--100 epochs, advantage temperature 1.0, BC coefficient 0.2--0.5, failure-penalty coefficient 0.5--1.0, and margin 4.0. Both use batch size 2048, learning rate $10^{-4}$, and value coefficient 50.0. Their endpoint comparison therefore combines differences in collection, optimization, update budget, and component selection.

\subsection{Paired Improvement Analysis}

Table~\ref{tab:go2_paired} reports a paired analysis of the Round~6 configuration ($\tau=0.5$) against the baseline. The analysis uses the 119,920 matched records identified by the common seed sequence. The transition counts describe proposal-sampled outcomes, whereas the reported target failure rates use importance weights.

Rounds 1--6 were evaluated with the preselected $\tau=0.8$. After the sixth update, validation rollouts on different scenarios or random seeds selected $\tau=0.5$ for the same specialist. On the reported test rollouts, the composed failure estimate is $2.43\times10^{-4}$ at $\tau=0.8$ and $1.47\times10^{-4}$ at $\tau=0.5$. The latter is a threshold change applied to the Round~6 model and the best observed setting among those tested.

\begin{table}[ht]
\centering
\small
\setlength{\tabcolsep}{2mm}
\begin{tabular}{@{}lcc@{}}
\toprule
\textbf{Transition} & \textbf{Count} & \textbf{Net} \\
\midrule
Failure $\to$ Success (improvement) & 681 & \\
Success $\to$ Failure (regression)  & 335 & \\
\midrule
\textbf{Net improvement} & & \textbf{+346 episodes} \\
\bottomrule
\end{tabular}
\caption{Go2 paired analysis: baseline vs.\ Round~6 ($\tau=0.5$).}
\label{tab:go2_paired}
\end{table}

\subsection{Terrain Parameter Mapping}

The 4-D terrain parameters are normalized to $[-1, 1]$ and discretized into 10 bins per dimension ($10^4 = 10{,}000$ candidates). Table~\ref{tab:go2_terrain} gives the mapping from normalized values to physical terrain parameters in MuJoCo.

\begin{table}[ht]
\centering
\small
\setlength{\tabcolsep}{3mm}
\begin{tabular}{@{}lccc@{}}
\toprule
\textbf{Dimension} & \textbf{Normalized} & \textbf{Physical Parameter} & \textbf{Physical Range} \\
\midrule
$cx$      & $[-1, 1]$ & Bump forward distance & $[0.1, 0.8]$\,m \\
$cy$      & $[-1, 1]$ & Bump lateral offset   & $[-0.30, 0.30]$\,m \\
$radius$  & $[-1, 1]$ & Bump radius            & $[0.1, 0.5]$\,m \\
$height$  & $[-1, 1]$ & Bump height            & $[-0.10, 0.10]$\,m \\
\bottomrule
\end{tabular}
\caption{Go2 normalized-to-physical terrain parameter mapping.}
\label{tab:go2_terrain}
\end{table}

In this coordinate system, the bump center $(cx, cy)$ is specified relative to the robot's current position in polar-like coordinates. Negative height values produce depressions rather than bumps.

\subsection{Failure Detection Criteria}

Table~\ref{tab:go2_failure} lists the failure detection conditions used in the MuJoCo environment. 

\begin{table}[ht]
\centering
\small
\setlength{\tabcolsep}{3mm}
\begin{tabular}{@{}lc@{}}
\toprule
\textbf{Event} & \textbf{Condition} \\
\midrule
Fallen          & Base height $< 0.15$\,m or tilt $> 1.3$\,rad  \\
Base collision  & Contact force $> 80$\,N                        \\
Thigh collision & Contact force $> 80$\,N                        \\
Stuck           & Speed $< 0.05$\,m/s                            \\
Timeout         & Terrain endpoint not reached within the horizon \\
Success         & Terrain endpoint reached within the horizon, without failure \\
\bottomrule
\end{tabular}
\caption{Go2 failure detection criteria. The recorded horizon is 40 steps for training and 30 for evaluation.}
\label{tab:go2_failure}
\end{table}

\section{ManiSkill Multi-Task Manipulation --- Additional Details}
\label{app:maniskill}

\subsection{Criticality Model}

The ManiSkill criticality model is a residual MLP with 512-unit hidden layers, ReLU activations, and skip connections. Its 51-D input comprises a 48-D observation, zero-padded across tasks, and a 3-D force vector. Its two-class logits (success / failure) supply a failure-class risk score. The model scores all $11^3 = 1331$ force-grid candidates per timestep. The criticality-guided sampling wrapper samples via $\varepsilon$-softmax ($\varepsilon = 0.01$) with inverse temperature $\beta = 3.0$ (Eq.~\ref{eq:q_rl} in the main text).

Training spans PushCube, PickCube, StackCube, and PegInsertionSide. We report the latter two because the recorded PushCube and PickCube results predate changes to their environment parameters and are not comparable to the current setup. The predictor is trained on per-step observation--force pairs with episode-level failure labels. Its train/validation/test split is 80/10/10, and successful examples are subsampled to 20\%. Training uses 30 epochs, Adam, learning rate $10^{-4}$, and batch size 4096, with checkpoint selection by validation AUC.

Policy updates use 50 epochs, batch size 1024, and a BC coefficient of 1.0. The learning rate is fixed at $10^{-5}$ in all rounds, with no value-only warmup. All rounds use PPO discount $\gamma = 0.95$, value coefficient 1.0, max gradient norm 1.0, and Adam optimizer ($\varepsilon = 10^{-5}$).

\subsection{Failure Indicator}

The ManiSkill evaluation records a binary terminal failure indicator, with $0$ on success and $1$ on failure. An episode terminates upon success or at the maximum horizon. Thus, failure is recorded when the task is not completed within the allotted steps.

\subsection{Criticality-Guided Sampling Algorithm}

The wrapper controls per-timestep force selection during data collection and evaluation. At each timestep, the wrapper applies the following rule.

\begin{enumerate}
\item \textbf{Grid construction}. The 3-D force space is discretized into an $11 \times 11 \times 11$ grid over $[-1, 1]^3$ (step 0.2), yielding 1331 candidate force vectors. For the two 3-D tasks reported here, this gives 1331 candidates. The 2-D PushCube task instead uses 121 force vectors with a zero-filled third coordinate.

\item \textbf{Risk scoring}. All 1331 candidates are scored by the criticality model $C_\phi$ in a single forward pass, producing a vector of risk scores $\mathbf{c} \in [0, 1]^{1331}$.

\item \textbf{Proposal distribution}. If $\max(\mathbf{c}) > 0.5$, the sampling distribution is:
\begin{equation}
q(\text{force}_i) = (1 - \varepsilon) \cdot \frac{\exp(3\mathbf{c}_i)}{\sum_j \exp(3\mathbf{c}_j)} + \frac{\varepsilon}{1331},
\end{equation}
with $\varepsilon = 0.01$. If $\max(\mathbf{c}) \leq 0.5$, uniform sampling is used because no candidate exceeds the score cutoff.

\item \textbf{Importance weight}. Each sampled force carries weight $w = p_{\text{unif}} / q(\text{force})$, where $p_{\text{unif}} = 1/1331$.

\item \textbf{Cumulative weight guard}. The product of per-step weights is tracked across the episode. If the cumulative weight drops below $10^{-2}$, the wrapper falls back to uniform sampling, so subsequent proposal ratios are one. The already accumulated ratio is retained.
\end{enumerate}

The same wrapper handles training data collection and evaluation. During evaluation, setting $\varepsilon = 1$ makes it sample uniformly for unbiased baseline estimation.

\section{LIBERO VLA Manipulation --- Additional Details}
\label{app:libero}

\subsection{Criticality Model}

A single, task-agnostic criticality MLP is trained on data from all 40 LIBERO tasks. Its architecture is a residual MLP with four Pre-LN blocks, 128-unit hidden layers, expansion factor 1, and Dropout 0.1. The input is a simulator initial-state vector right-padded to 92-D. The output is a risk score. Training uses binary cross-entropy on rollout outcomes collected from the warmup-adapted baseline on an expanded initial-state pool (40,000 self-generated scenes, 1000 per task, disjoint from the 50 official evaluation states per task). This predictor pool is separate from the 2000 official training demonstrations used for BC; those demonstrations are also disjoint from the evaluation states.

\subsection{Gate Threshold Selection}

Table~\ref{tab:libero_gate} reports the fixed per-suite thresholds used in the evaluated composed policy and their corresponding failure rates.

\begin{table}[ht]
\centering
\small
\setlength{\tabcolsep}{3mm}
\begin{tabular}{@{}lccccc@{}}
\toprule
\textbf{Suite} & $\boldsymbol{\tau}$ & \textbf{Specialist only} $\downarrow$ & \textbf{Baseline} $\downarrow$ & \textbf{Ours} $\downarrow$ & \textbf{vs.\ Baseline} $\downarrow$ \\
\midrule
LIBERO-10      & 0.9 & 4.20\% & 2.93\% & 2.20\% & $-$25.0\% \\
LIBERO-Goal    & 0   & 1.40\% & 1.87\% & 1.40\% & $-$25.0\% \\
LIBERO-Object  & 0.5 & 1.00\% & 0.53\% & 0.27\% & $-$50.0\% \\
LIBERO-Spatial & 0.6 & 1.93\% & 2.13\% & 1.73\% & $-$18.8\% \\
\midrule
Average        & --- & 2.13\% & 1.87\% & 1.40\% & $-$25.0\% \\
\bottomrule
\end{tabular}
\caption{LIBERO per-suite gate thresholds and component-only diagnostics. The average combines four suites with different thresholds. Relative changes use unrounded failure rates.}
\label{tab:libero_gate}
\end{table}

LIBERO-Goal uses $\tau = 0$, assigning all evaluated episodes to the specialist. Within that suite, its failure rate is lower than the nominal policy's. For the other suites, the gate dispatches high-criticality episodes to the specialist and retains the frozen nominal policy for nominal cases.

\subsection{Evaluation Protocol}

All models are evaluated on the 50 official initial states per task (2000 total episodes across 40 tasks). Each model is evaluated in 3 independent runs, with per-run randomization coming from environment nondeterminism. The main text reports the average failure rate across the 3 runs. 

\subsection{Fine-Tuning Setups}
Both methods fine-tune from the same 40K-step warmup checkpoint for 40,000 continuation iterations with matched hyperparameters (AdamW, lr=$5\times10^{-6}$, batch size 32, cosine decay, VLM backbone frozen). For specialist training, we select 1000 demonstration slots from the 2000 official LIBERO training demonstrations with proposal $q(i)$ and draw BC training trajectories from the selected set with normalized $p(i)/q(i)$ weights. For fine-tuning on randomly collected data, we select the same number uniformly, giving constant correction weights. The BC loss itself is unweighted.

\section{RoboTwin Bimanual Manipulation --- Detailed Results}
\label{app:robotwin}

\subsection{Criticality Model}

The RoboTwin criticality model is a global state-only MLP with the following configuration.

\begin{itemize}
\item Its architecture is LayerNorm(437) $\to$ Linear(437, 512) $\to$ GELU $\to$ Dropout(0.15) $\to$ Linear(512, 512) $\to$ GELU $\to$ Dropout(0.15) $\to$ Linear(512, 256) $\to$ GELU $\to$ Dropout(0.15) $\to$ Linear(256, 128) $\to$ GELU $\to$ Linear(128, 1).
\item Its input is a flattened 437-D object-level initial scene state vector (no task ID).
\item The network emits a scalar logit. Its sigmoid gives the risk score $C_\phi(s_0) = \sigma(\text{logit}) \in [0, 1]$ used for sampling and routing.
\item Training uses 200 epochs, AdamW (lr=$10^{-3}$, weight decay=$10^{-4}$), batch size 256, binary cross-entropy with class-balanced pos\_weight. Best checkpoint selected by validation AP.
\end{itemize}

The dataset contains 5003 rollout episodes (70/15/15 train/val/test split). Test-set metrics are AP 0.902, ROC AUC 0.942, accuracy 0.888 (threshold 0.5).

\subsection{Importance Sampling Data Construction}

The evaluated global proposal mixes the original candidate-pool mass with
normalized risk scores:

\begin{equation}
q(i)=0.1p(i)+0.9\frac{C_\phi(s_0^{(i)})}{\sum_j C_\phi(s_0^{(j)})},
\end{equation}

\noindent where $p(i)$ is the original distribution over the full candidate pool.
There is no task-balanced bucket allocation or separate nearest-neighbor
failure-density term in the evaluated global-$q$ run. The selected subset has
5000 slots representing 2574 unique source episodes.

For the specialist, each selected slot retains its $p(i)/q(i)$ weight. We sum weights for repeated source episodes and resample trajectories with replacement using normalized totals. The BC loss remains unweighted. Both fine-tuning methods use a 90/10 mixture with 90\% from the selected subset and 10\% full replay from the original RoboTwin2 training data. Fine-tuning uses action-head-only BC (VLM backbone frozen), AdamW, lr=$6\times10^{-5}$, cosine schedule with 500-step warmup, batch size 32, 40,000 steps. All training seeds are disjoint from evaluation seeds.

\subsection{Gate Diagnostics}

The evaluated system uses a single global gate threshold $\tau=0.65$, selected on the validation scenes used to train the criticality model and held fixed for the common 1000-episode test manifest. Both scenario scoring and component selection use the initial state $s_0$. The standalone specialist has 16.30\% aggregate failure, versus 14.40\% for the nominal policy and 13.30\% for the returned composed policy. This component-level comparison shows a nominal/critical trade-off similar to LIBERO (Table~\ref{tab:libero_gate}). The available endpoint results do not quantify how stable $0.65$ is across independent calibration sets.

At $\tau = 0.65$, the gate selects the specialist on 124 of 1000 evaluation episodes and the nominal policy on the remaining 876. Nineteen nominal-policy failures become successes, while eight nominal-policy successes regress. The net gain of 11 episodes gives the composed policy a 13.30\% failure rate, a 7.6\% relative reduction on this manifest.

\subsection{Seed Disjointness Audit}

Training candidates exclude the evaluation seed window (2,030,000--2,079,019). The reported global-proposal subset contains 5000 sampled slots and 2574 unique source episodes. Fine-tuning on randomly collected data uses 5000 uniformly selected training episodes. These training partitions are separate from the common 1000-episode evaluation manifest.

\section{Real-Robot Banana-on-Plate --- Additional Details}
\label{app:real}

\subsection{Criticality Pipeline}

The criticality model uses 8-D geometric coordinates extracted from the initial RGB frame:

\begin{itemize}
\item Hue-saturation-value (HSV) segmentation uses plate (H $\in$ [43, 88], S $\in$ [28, 185], V $\in$ [35, 205]) and banana (H $\in$ [8, 38], S $\in$ [65, 255], V $\in$ [45, 255]).
\item Landmark extraction uses principal component analysis (PCA) endpoint analysis with dark-patch centroid method for the banana stem and opposite endpoint.
\item The coordinate extractor outputs an 8-D vector [plate\_x, plate\_y, banana\_x, banana\_y, stem\_x, stem\_y, opposite\_x, opposite\_y].
\end{itemize}

The criticality model is a two-layer MLP (8$\to$32$\to$1, ReLU, Dropout 0.5) that outputs a risk score. It is trained for 500 epochs (lr=$3\times10^{-3}$, weight decay=$6\times10^{-3}$, batch size 16, early stopping patience 80). The recorded fit to 99 labeled episodes has AUC 0.918, F1 0.821, and accuracy 0.859. The recorded training configuration uses no held-out validation split, so these metrics do not establish out-of-sample predictor performance.

\subsection{Importance Sampling and Results}

Criticality-guided collection scores 10,000 randomly generated candidate scenarios, then selects 100 via $\varepsilon$-mixed sampling ($\varepsilon=0.1$). Risk scores are normalized over the candidate pool before mixing with the uniform distribution, as in Equation~\ref{eq:q_il}. BC training draws trajectories from this selected set with normalized $p(i)/q(i)$ weights. Failed rollouts are used only for criticality-model fine-tuning, never as BC supervision. The gate uses the 8-D coordinates extracted from the initial RGB frame.

Initial training uses 100 episodes collected under the default random-collection
protocol. The two continuation datasets contain 100 episodes each, collected
using uniform sampling for Random and criticality-guided sampling for the
specialist. Both continuations start from the same checkpoint and use 10,000
fine-tuning steps.

The gate threshold is selected on 100 initial-state validation scenes from the
same collection pool used for criticality fitting, each executed once with the nominal policy and once
with the specialist. Candidate thresholds were swept on a 0.05 grid by routed
failure rate, and $\tau=0.15$ was retained; ties in the sweep favour the setting
that sends the fewest scenes to the specialist. None of those validation scenes
appears among the 50 reported test scenes. The selection protocol is implemented
in \texttt{select\_threshold.py}.

Across 50 paired trials, the recorded proposal ratios and binary outcomes cover
the nominal policy, the policy fine-tuned on randomly collected data, and the specialist. Applying
$\tau=0.15$ to the recorded risk scores yields the reported 3.72\% composed
estimate. Each bootstrap replicate resamples complete paired trials. The
resulting intervals and paired differences appear in Tables~\ref{tab:real} and~\ref{tab:paired_gate}.

For comparison, the standalone specialist has 6.05\% estimated failure (95\% CI
1.83--11.93\%), versus 11.15\% for the nominal policy and 3.72\% for the
composed policy. Its paired failure-rate difference from the composed policy is
2.33 percentage points (95\% bootstrap CI 0--6.99).

\section{Criticality Model Architecture Summary}
\label{app:criticality_summary}

Table~\ref{tab:criticality_summary} compares the criticality models across domains. Although they range from 321 to 1.08M parameters and use different state representations, all are trained with cross-entropy on binary rollout labels. Model-selection procedures are domain-specific.

\begin{table}[ht]
\centering
\small
\setlength{\tabcolsep}{1.2mm}
\begin{tabular}{@{}lccccc@{}}
\toprule
\textbf{Domain} & \textbf{Regime} & \textbf{Architecture} & \textbf{Input Dim} & \textbf{Params} & \textbf{Candidate set} \\
\midrule
Go2         & RL & 3-layer MLP (56$\to$256$\to$256$\to$2) & 56 & 0.081M & 10,000 \\
ManiSkill   & RL & ResMLP (51$\to$512$\to$2)    & 51 & 1.078M & 1,331 \\
LIBERO      & BC & 4-block ResMLP (92$\to$128)  & 92 & 0.146M & 2,000 demos \\
RoboTwin    & BC & 5-layer MLP (437$\to$512$\to$...$\to$1) & 437 & 0.652M & Demo pool \\
Real robot  & BC & 2-layer MLP (8$\to$32$\to$1)         & 8 & 0.0003M & 10,000 \\
\bottomrule
\end{tabular}
\caption{Criticality model architecture summary. RL identifies the RL-pretrained systems; BC identifies the three expert-demonstration continuations. LIBERO and RoboTwin use initial-state criticality (one score per episode) rather than per-timestep scoring. Counts refer to the specialist-data candidate pool or perturbation grid, not selected training slots. The separate LIBERO predictor pool has 40,000 self-generated scenes.}
\label{tab:criticality_summary}
\end{table}

\section{Criticality Model Inference Benchmarks}
\label{app:time}

We benchmark criticality-model forward-pass latency on a single NVIDIA RTX 4090 GPU (PyTorch 2.6, CUDA 12.4). Each model is tested at the candidate batch size listed in Table~\ref{tab:time}. The measurements exclude input preparation, transfer, and policy-routing overhead.

\begin{table}[ht]
\centering
\small
\setlength{\tabcolsep}{1.5mm}
\begin{tabular}{@{}lcccc@{}}
\toprule
\textbf{Experiment} & \textbf{Params} & \textbf{Batch} & \textbf{Latency (ms)} & \textbf{Throughput (samples/s)} \\
\midrule
Go2 MuJoCo        & 0.081M  & 10,000 & 0.183 $\pm$ 0.002 & $5.45{\times}10^7$ \\
ManiSkill    & 1.078M  & 1,331  & 0.345 $\pm$ 0.005 & $3.86{\times}10^6$ \\
Real robot        & 0.0003M & 10,000 & 0.131 $\pm$ 0.003 & $7.63{\times}10^7$ \\
RoboTwin          & 0.652M  & 10,000 & 0.468 $\pm$ 0.004 & $2.14{\times}10^7$ \\
LIBERO            & 0.146M  & 10,000 & 0.475 $\pm$ 0.005 & $2.11{\times}10^7$ \\
\bottomrule
\end{tabular}
\caption{Criticality model inference latency. Measured on a single NVIDIA RTX 4090, PyTorch 2.6, CUDA 12.4. Mean $\pm$ std over 200 timed iterations after 50 warmup iterations.}
\label{tab:time}
\end{table}

The largest criticality model (ManiSkill, 1.08M parameters) scores 1331 force-grid candidates in 0.35 ms. The Go2 and real-robot models process 10,000 candidates in under 0.2 ms. The largest measured forward-pass latency, 0.475 ms, is 2.4\% of a 20 ms control period on this hardware. These timings do not measure end-to-end control latency.

\section{Full Round-by-Round Results}
\label{app:rounds}

\begin{figure}[ht]
\centering
\includegraphics[width=0.88\textwidth]{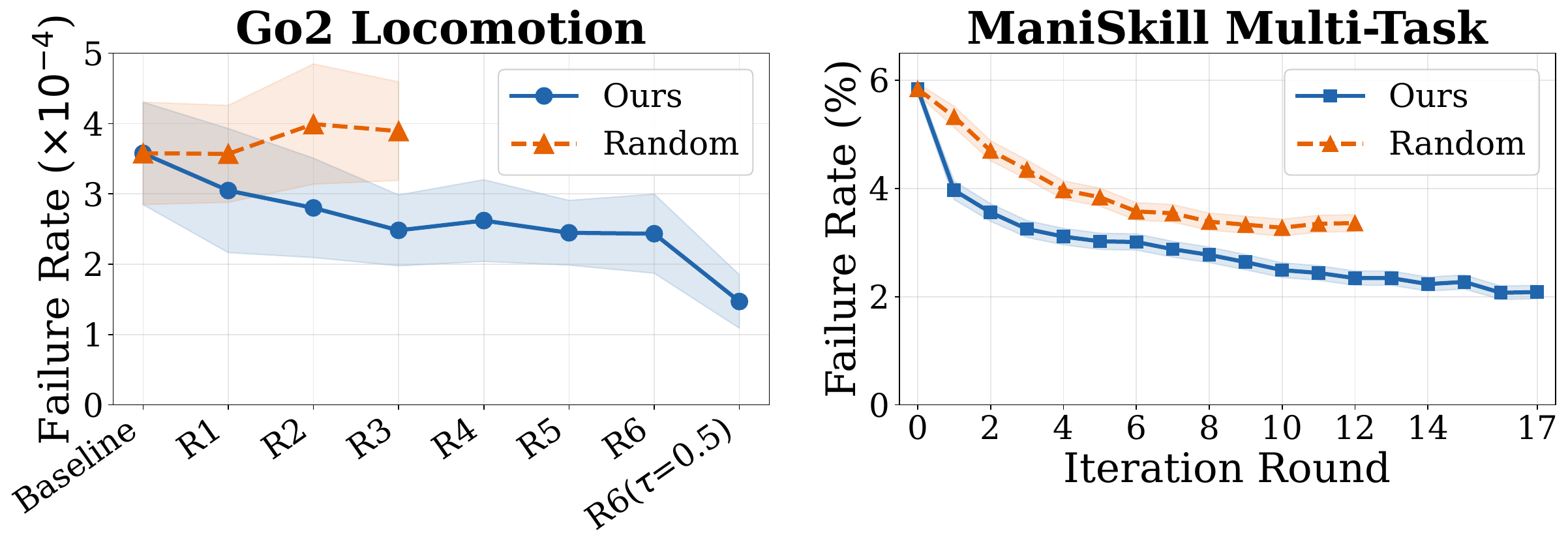}
\caption{Complete multi-round failure trends with 95\% intervals: bootstrap intervals for Go2 and binomial normal approximations for the two-task ManiSkill average. Go2's last point changes the Round~6 gate threshold from 0.8 to 0.5 without another training round. The main-text figure shows means only.}
\label{fig:trend_intervals}
\end{figure}

Tables~\ref{tab:go2_full} and~\ref{tab:maniskill_full} report all recorded rounds. The common budgets are the shorter observed runs: three rounds for Go2 and 12 for ManiSkill. Within these windows, each arm's best value is its lowest observed test failure rate. These descriptive minima are susceptible to selection bias; no independent validation-based round selection is established by the available records. We therefore also report the exact common-round comparisons and Ours' last recorded rounds (6 and 17). Gate selection is separate: its validation protocol is described in Appendix~\ref{app:thresholds}. Equal rounds do not match total acquisition or compute cost, and Go2 additionally differs in optimization settings.

\begin{table}[ht]
\centering
\small
\setlength{\tabcolsep}{0.5mm}
\begin{tabular}{@{}lcc@{}}
\toprule
\textbf{Configuration} & \textbf{Failure Rate} $\downarrow$ & \textbf{vs.\ Baseline} $\downarrow$ \\
\midrule
Baseline                & $3.58 \times 10^{-4}$ & --- \\
\midrule
\multicolumn{3}{@{}l}{\emph{Ours (criticality-guided collection)}} \\
Round 1                               & $3.05 \times 10^{-4}$ & $-$14.8\% \\
Round 2                               & $2.80 \times 10^{-4}$ & $-$21.7\% \\
Round 3                               & $2.48 \times 10^{-4}$ & $-$30.6\% \\
Round 4                               & $2.62 \times 10^{-4}$ & $-$26.7\% \\
Round 5                               & $2.45 \times 10^{-4}$ & $-$31.6\% \\
Round 6 ($\tau = 0.8$)                & $2.43 \times 10^{-4}$ & $-$32.0\% \\
\textbf{Round 6 ($\tau = 0.5$)}       & $\mathbf{1.47 \times 10^{-4}}$ & \textbf{$-$58.9\%} \\
\midrule
\multicolumn{3}{@{}l}{\emph{Fine-tuning on randomly collected data (Random)}} \\
Round 1                               & $3.57 \times 10^{-4}$ & $-$0.2\% \\
Round 2                               & $3.99 \times 10^{-4}$ & $+$11.6\% \\
Round 3                               & $3.89 \times 10^{-4}$ & $+$8.8\% \\
\bottomrule
\end{tabular}
\caption{Go2 full round-by-round results. 119,920--120,000 completed importance-sampled test episodes per configuration. Rounds 1--6 use the preselected $\tau=0.8$. The final $\tau=0.5$ row evaluates the same Round~6 specialist after separate threshold validation.}
\label{tab:go2_full}
\end{table}

\begin{table}[ht]
\centering
\footnotesize
\setlength{\tabcolsep}{2.5mm}
\begin{tabular}{@{}lccc@{}}
\toprule
\textbf{Configuration} & \textbf{StackCube} $\downarrow$ & \textbf{PegInsertionSide} $\downarrow$ & \textbf{Avg.} $\downarrow$ \\
\midrule
Baseline                   &   5.01 &   6.67 &   5.84 \\
\midrule
\multicolumn{4}{@{}l}{\emph{Ours (criticality-guided collection)}} \\
Round 1                    &   3.80 &   4.13 &   3.97 \\
Round 2                    &   3.24 &   3.88 &   3.56 \\
Round 3                    &   2.99 &   3.51 &   3.25 \\
Round 4                    &   2.86 &   3.36 &   3.11 \\
Round 5                    &   2.80 &   3.25 &   3.02 \\
Round 6                    &   2.74 &   3.28 &   3.01 \\
Round 7                    &   2.56 &   3.19 &   2.88 \\
Round 8                    &   2.53 &   3.02 &   2.77 \\
Round 9                    &   2.19 &   3.08 &   2.64 \\
Round 10                   &   2.02 &   2.96 &   2.49 \\
Round 11                   &   1.89 &   2.99 &   2.44 \\
Round 12                   &   1.77 &   2.91 &   2.34 \\
Round 13                   &   1.80 &   2.88 &   2.34 \\
Round 14                   &   1.65 &   2.82 &   2.23 \\
Round 15                   &   1.57 &   2.97 &   2.27 \\
\textbf{Round 16 (best)} & \textbf{1.48} & \textbf{2.66} & \textbf{2.07} \\
Round 17 (final) & 1.48 & 2.68 & 2.08 \\
\midrule
\multicolumn{4}{@{}l}{\emph{Fine-tuning on randomly collected data (Random)}} \\
Round 1                    &   4.50 &   6.15 &   5.32 \\
Round 2                    &   4.01 &   5.38 &   4.70 \\
Round 3                    &   3.35 &   5.34 &   4.35 \\
Round 4                    &   3.02 &   4.92 &   3.97 \\
Round 5                    &   2.89 &   4.79 &   3.84 \\
Round 6                    &   2.72 &   4.43 &   3.58 \\
Round 7                    &   2.45 &   4.63 &   3.54 \\
Round 8                    &   2.21 &   4.56 &   3.39 \\
Round 9                    &   2.10 &   4.56 &   3.33 \\
\textbf{Round 10 (best)} & \textbf{1.88} & \textbf{4.66} & \textbf{3.27} \\
Round 11                   &   1.92 &   4.77 &   3.35 \\
Round 12 (final) & 1.94 & 4.77 & 3.36 \\
\bottomrule
\end{tabular}
\caption{ManiSkill full round-by-round results. Failure rate (\%). Approximately 100,000 baseline episodes per task and 25{,}000 per task per update round. The preselected $\tau=0$ remains the reported setting after training. Criticality-guided training and fine-tuning on randomly collected data run for 17 and 12 rounds, respectively.}
\label{tab:maniskill_full}
\end{table}

\section{Compute Infrastructure}
\label{app:compute}

Table~\ref{tab:compute} lists the full hardware configuration for each experiment group. All training and evaluation were performed on Linux clusters.

\begin{table}[ht]
\centering
\resizebox{\textwidth}{!}{%
\begin{tabular}{@{}llll@{}}
\toprule
\textbf{Experiment Group} & \textbf{CPU} & \textbf{Logical Cores} & \textbf{GPU} \\
\midrule
LIBERO                   & 2$\times$ Intel Xeon Platinum 8369B @ 2.90\,GHz & 128 & 8$\times$ NVIDIA A100-SXM4-80GB \\
RoboTwin (specialist / Random)       & Linux cluster & --- & NVIDIA A100 (16 / 4 GPUs) \\
Real robot (train)        & 2$\times$ Intel Xeon Platinum 8369B @ 2.90\,GHz & 128 & 8$\times$ NVIDIA A100-SXM4-80GB \\
Real robot (test)         & Intel Core Ultra 7 270K Plus & 24 & 2$\times$ NVIDIA GeForce RTX 5090 \\
Go2, ManiSkill            & 2$\times$ Intel Xeon Platinum 8336C @ 2.30\,GHz & 128 & 2$\times$ NVIDIA GeForce RTX 4090 (24\,GB) \\
\bottomrule
\end{tabular}}
\caption{Compute infrastructure per experiment group.}
\label{tab:compute}
\end{table}

LIBERO training used all 8 GPUs via Accelerate Distributed Data Parallel (DDP) with bfloat16 mixed precision. The RoboTwin specialist used 16 GPUs and fine-tuning on randomly collected data used four. Both used global batch size 32. Real-robot training used the 8$\times$A100 configuration, while real-robot inference ran on a separate machine with two RTX 5090 GPUs. Go2 and ManiSkill experiments ran on a dual-socket machine with two RTX 4090 GPUs. The inference benchmarks in Appendix~\ref{app:time} used a single RTX 4090 from this machine. All criticality models fit in GPU memory alongside the policy, requiring no dedicated inference hardware.

\section{Random + gate Comparisons and Paired Uncertainty}
\label{app:random_gate}

\subsection{Configuration and validation selection}
Random denotes fine-tuning on randomly collected data. Random + gate replaces the specialist trained with criticality-guided data in Equation~\ref{eq:gate} with that policy, retaining the nominal policy, criticality scores, routing rule, and evaluation manifest. Each arm's reported threshold was chosen on validation rollouts and then held fixed for testing. The 0.05-grid scans of the test manifests reproduce these fixed configurations; they are not the source of the reported threshold choices. For LIBERO and RoboTwin, both arms use 21 candidates from 0 to 1; the same per-suite or global selection granularity applies to both arms. The physical comparison uses the validation-selected 0.15 for both.

Table~\ref{tab:random_gate_thresholds} gives the selected values; Appendix~\ref{app:thresholds} specifies the routing conventions.

\subsection{Holding the gate fixed}
Using Ours' thresholds for both variants yields Random + gate failure rates of 1.72\% for LIBERO, 14.10\% for RoboTwin, and 8.53\% for the physical robot, compared with 1.40\%, 13.30\%, and 3.72\% for Ours. Allowing Random its own validation-selected thresholds gives 1.68\%, 14.00\%, and 8.53\%. These checks distinguish the trained components under the same gating interface. They do not identify separate effects of collection and inverse-proposal resampling.

For Go2, Random + gate uses the best observed Random checkpoint (Round~1). Its updated failure estimate is $3.38\times10^{-4}$ at validation-selected $\tau=0.35$; no interval is available for this updated configuration.

For RoboTwin at $\tau=0.60$, Random + gate routes 137 episodes, repairs 19 baseline failures, and introduces 15 regressions. At $\tau=0.65$, it routes the same 124 episodes as Ours, repairing 16 and introducing 13, versus 19 and eight for Ours. For the physical task, both arms route 42 of 50 trials to their fine-tuned component and eight to the nominal policy.

An improvement from Baseline to Random + gate includes both fine-tuning on randomly collected data and composition. Dividing it by the improvement from Baseline to Ours is a descriptive ratio, not a causal percentage attributable to gating. The direct within-component comparison is Random versus Random + gate, or Specialist only versus Ours.

\subsection{Paired comparisons}
For LIBERO, each of the 2,000 initial scenes has three recorded evaluations per policy. We use the raw turn outcomes, average within each scene, and resample scenes within each task while preserving all policies and repetitions as a paired block. Thus, the intervals do not treat 6,000 executions as 6,000 distinct starts. As a sensitivity check, resampling complete tasks within each of the four suites preserves within-task dependence. Each analysis uses 100,000 percentile-bootstrap replicates. Physical comparisons instead resample complete paired trials, keeping the original importance weights and gate decisions.

\begin{table}[ht]
\centering
\small
\setlength{\tabcolsep}{2mm}
\begin{tabular}{@{}llcc@{}}
\toprule
Domain & Failure-rate difference & Estimate (pp) & 95\% interval (pp) \\
\midrule
LIBERO & Baseline $-$ Ours & 0.47 & $[0.12,0.82]$ \\
LIBERO & Random + gate $-$ Ours & 0.28 & $[-0.10,0.67]$ \\
LIBERO & Random $-$ (Random + gate) & 0.13 & $[-0.05,0.33]$ \\
LIBERO & Specialist only $-$ Ours & 0.73 & $[0.42,1.07]$ \\
Real robot & Random + gate $-$ Ours & 4.81 & $[0.00,9.90]$ \\
Real robot & Random $-$ (Random + gate) & 9.52 & $[1.97,19.59]$ \\
\bottomrule
\end{tabular}
\caption{Paired failure-rate differences in percentage points (pp). Positive differences favor the policy to the right of the minus sign. LIBERO intervals resample initial scenes within task; physical intervals use weighted paired trials. Values are conditional on fixed trained policies and validation-selected thresholds.}
\label{tab:paired_gate}
\end{table}

LIBERO's task-cluster sensitivity intervals are $[0.07,0.98]$ pp for Baseline minus Ours and $[-0.15,0.87]$ pp for Random + gate minus Ours. The latter includes zero under either resampling scheme. For the physical comparison, the lower bound rounds to zero and the weight effective sample size is 12.44, so the gated difference remains imprecise. These intervals exclude uncertainty from training seeds, predictor fitting, threshold selection, and choosing a round from a learning curve.

For RoboTwin, the reported 19 improvements and eight regressions imply the two-sided exact McNemar value $2\sum_{k=0}^{8}\binom{27}{k}2^{-27}=0.05224$, under independent episode pairs. This calculation uses the published paired counts and does not model task clustering. Additional training seeds and calibration splits have not been evaluated here.

\subsection{Attribution not resolved by the present comparisons}
There is no matched experiment removing inverse-proposal resampling or freezing the initial criticality model throughout the multi-round updates. The reported results therefore measure complete procedures. Risk-score correlations with BC gradient norms or repairability, and robot-policy gradient correlations in Proposition~\ref{prop:rho}, have not been measured. The coverage toy study changes both critical and nominal coverage and does not validate the complete learned sampling pipeline. For RoboTwin, 5,000 selected slots contain 2,574 unique episodes (51.5\% unique; 48.5\% repeated slots), which describes the selected buffer but does not itself establish a variance reduction. Training-buffer weight effective sample sizes are distinct from the physical evaluation ESS and are not reported here.

\section{Data Acquisition and Deployment Costs}
\label{app:cost_ledger}

\begin{table}[ht]
\centering
\small
\setlength{\tabcolsep}{1.5mm}
\begin{tabular}{@{}lccc@{}}
\toprule
Domain & Predictor data & Continuation data per arm & BC updates per arm \\
\midrule
LIBERO & 40,000 scenes & 1,000 slots from 2,000 demos & 40,000 \\
RoboTwin & 5,003 rollouts & 5,000 slots + 10\% full replay & 40,000 \\
Real robot & 99 labeled episodes & 100 expert demonstrations & 10,000 \\
\bottomrule
\end{tabular}
\caption{Known acquisition and fine-tuning quantities. Predictor data are additional to the BC demonstration subsets. Some calibration scenes are reused from the predictor data, but evaluating each candidate component still requires rollout outcomes. These counts are not end-to-end wall-clock or GPU-hour measurements.}
\label{tab:cost_ledger}
\end{table}

LIBERO's common warmup uses all 2,000 official demonstrations for 40,000 updates before the matched continuations. Its predictor's 40,000 generated scenes are a separate data-collection cost. RoboTwin uses existing expert demonstrations and 5,003 predictor rollouts. The physical baseline is first fine-tuned on 100 expert demonstrations, followed by 100 additional demonstrations for each continuation. Demonstration acquisition time, total predictor-collection time, and the full calibration cost are not available as a complete ledger. We consequently make no claim of lower total acquisition or compute cost from matching BC steps alone.

The LIBERO and RoboTwin gates require simulator state beyond the action policies' RGB observations. The physical coordinate extractor supplies a different, task-specific interface. Reported criticality-model forward-pass timings exclude input extraction and transfers, routing, action-policy execution, and the memory footprint of retaining both policies. End-to-end deployment latency and peak memory remain unmeasured.

\end{document}